\documentclass[9pt,journal]{IEEEtran}
\IEEEoverridecommandlockouts

\usepackage[noadjust]{cite}  
\usepackage{amsmath,amssymb,amsfonts}
\usepackage{graphicx}
\usepackage{textcomp}
\usepackage{xcolor}
\usepackage{enumitem}
\usepackage[linesnumbered,ruled,vlined]{algorithm2e}
\usepackage{booktabs}

\usepackage{url}  
\usepackage{hyperref}  

\usepackage{array}  

\ifCLASSOPTIONcompsoc
    \usepackage[caption=false,font=normalsize,labelfont=sf,textfont=sf]{subfig}
\else
    \usepackage[caption=false,font=footnotesize]{subfig}
\fi

\usepackage{algpseudocode}  

\usepackage{nomencl}
\makenomenclature

\usepackage{etoolbox}

\usepackage{pifont}  

\SetCommentSty{mycommfont}

\SetKwInput{KwInput}{Input}  
\SetKwInput{KwOutput}{Output}  

\usepackage{algorithm}

\def\BibTeX{{\rm B\kern-.05em{\sc i\kern-.025em b}\kern-.08em
    T\kern-.1667em\lower.7ex\hbox{E}\kern-.125emX}}

\begin{document}

\title{Adaptive BESS and Grid Setpoints Optimization: A Model-Free Framework for Efficient Battery Management under Dynamic Tariff Pricing} 

\author{
    Alaa Selim\textsuperscript{*}, 
    Huadong Mo\textsuperscript{†}, 
    Hemanshu Pota\textsuperscript{*}, 
    Daoyi Dong\textsuperscript{‡}
    
    \thanks{\textsuperscript{*}School of Engineering and Information Technology, University of New South Wales, Canberra, Australia. e-mails: a.selim@unsw.edu.au, h.pota@unsw.edu.au}
    \thanks{\textsuperscript{†}School of Systems and Computing, University of New South Wales, Canberra, Australia. e-mail: huadong.mo@unsw.edu.au}
    \thanks{\textsuperscript{‡}CIICADA Lab, School of Engineering, Australian National University, ACT, Australia. e-mail: daoyi.dong@anu.edu.au}
    \thanks{Research work affiliated with the School of Engineering and Information Technology, University of New South Wales, Canberra, Australia.}
}

\maketitle

\begin{abstract}
This paper introduces an enhanced framework for managing Battery Energy Storage Systems (BESS) in residential communities. The non-convex BESS control problem is first addressed using a gradient-based optimizer, providing a benchmark solution. Subsequently, the problem is tackled using multiple Deep Reinforcement Learning (DRL) agents, with a specific emphasis on the off-policy Soft Actor-Critic (SAC) algorithm. This version of SAC incorporates reward refinement based on this non-convex problem, applying logarithmic scaling to enhance convergence rates. Additionally, a safety mechanism selects only feasible actions from the action space, aimed at improving the learning curve, accelerating convergence, and reducing computation times. Moreover, the state representation of this DRL approach now includes uncertainties quantified in the entropy term, enhancing the model's adaptability across various entropy types. This developed system adheres to strict limits on the battery's State of Charge (SOC), thus preventing breaches of SOC boundaries and extending the battery lifespan. The robustness of the model is validated across several Australian states' districts, each characterized by unique uncertainty distributions. By implementing the refined SAC, the SOC consistently surpasses 50 percent by the end of each day, enabling the BESS control to start smoothly for the next day with some reserve. Finally, this proposed DRL method achieves a mean reduction in optimization time by 50 percent and an average cost saving of 40 percent compared to the gradient-based optimization benchmark. 

\end{abstract}


\section{Introduction}

\IEEEPARstart{R}{ecently}, the deployment of renewable energy sources like photovoltaic (PV) generation has seen a significant increase, driven by the urgent need to mitigate the environmental impact of conventional energy sources \cite{1}. An integral component of these renewable systems is energy storage, often in the form of batteries, which helps manage the intermittency of renewable generation and meet varying energy demands \cite{2}. A common challenge with these systems is the optimal energy management to ensure efficient utilization of the generated energy and the battery storage \cite{3}. This involves balancing the power generation and demands, while considering factors such as grid tariffs, energy storage capacity, and the state of charge (SOC) of the battery \cite{4} and \cite{5}.


Historically, various model-based control algorithms have been employed for managing the energy flow in Battery Energy Storage System (BESS). These algorithms range from Model Predictive Control (MPC) \cite{9} to Linear Quadratic Regulator (LQR) \cite{10} and Nonlinear Programming (NLP) \cite{11}. For example, Parisio et al. \cite{12} developed an MPC approach to optimize microgrid operations, inclusive of BESS. Although their work achieved significant results, the approach struggled with the non-convex nature of BESS problems. This non-convexity arises due to the complex interactions between the battery's charging and discharging processes, state of charge dynamics, and the inclusion of operational constraints like power limits and efficiency losses. These factors introduce non-linear constraints and bilinear terms.  Likewise, the work of Li and Wang \cite{13} using LQR, and that of Belotti and Pietro \cite{14} utilizing NLP, encountered similar challenges when dealing with non-convex problems, such as difficulties in achieving global optimality and sensitivity to model inaccuracies and external disturbances.

Additionally, while the classical BESS control problem, as outlined in \cite{wu2017optimal} and \cite{wu2017optimal2}, was typically addressed using convex formulations, challenges arose when variable penalty terms and non-linear constraints were introduced. This resulted in Disciplined Convex Programming (DCP) issues \cite{grant2006disciplined}. DCP provides guidelines to ensure that problem formulations remain convex, facilitating their solution with certain optimization solvers. However, the inclusion of these terms shifted the problem into the non-convex domain, posing challenges for conventional solvers.

An exploration of the literature unveils several potential solutions for non-convex problems: 
Swarm Intelligence (SI) algorithms, inspired by the collective behavior of social organisms, have also been effectively used in the global optimization of non-convex problems \cite{32}. Dorigo et al. \cite{32} demonstrated the application of Ant Colony Optimization, a popular SI technique, in finding near-global solutions for non-convex problems. Despite their advantages, SI algorithms depend on careful parameter tuning and can become trapped in local optima. Semi-definite Programming (SDP) is another advanced mathematical optimization method that has been employed to solve non-convex problems \cite{24}. Zhou et al. \cite{25} demonstrated that SDP could provide global solutions under certain conditions, but they also noted its limitations, such as the requirement for specific problem structures and the inability to handle integer variables. The Homotopy method is another technique aimed at finding global optima in non-convex optimization problems \cite{26}. Zhang et al. \cite{26} showed that by employing a continuous deformation of the cost function, they could trace the optimal solutions of the deformed problem to find the global minimum of the original problem. While this method has shown potential in achieving global optimality, it is not immune to challenges, particularly when dealing with complex, high-dimensional problems.

Given these limitations of model-based and heuristic algorithms, there has been a growing interest in model-free algorithms. 
These algorithms, which are not dependent on an accurate model of the system, show a greater capability to adapt to unexpected environmental changes. Addressing the intricacies of non-convex optimization, two methods, the Adam optimizer and Deep Reinforcement Learning (DRL), have demonstrated noteworthy efficacy in \cite{15,16,17,18,19,20,21}. The Adam optimizer \cite{Adam}, a gradient-driven technique, amalgamates the benefits of AdaGrad \cite{adagrad} and RMSProp \cite{rmsprop} algorithms. Especially adept at navigating large-scale problems replete with noisy gradients, its adaptive learning rates often lead to rapid convergence. This makes Adam particularly valuable for benchmarking feasible solutions in non-convex settings. However, its susceptibility to local minima or saddle points can sometimes limit its exploration scope. Contrarily, DRL offers a more expansive search method \cite{wei2021deep} and \cite{ren2018self}. Unconstrained by traditional optimization paradigms, DRL learns interactively, providing the potential to surpass local optima and uncover more optimal solutions. While its model-free nature enhances its versatility, the success of DRL is contingent upon precise reward function design, architecture decisions, and hyperparameter choices. In this study, the efficacy and adaptability of both the Adam optimizer and DRL are thoroughly investigated for our problem formulation.

In the context of day-ahead scheduling problems for battery and grid control in smart buildings, previous research has encountered numerous limitations in their quest for optimal solutions. For instance, studies such as \cite{koskela2019using}, \cite{yao2023two}, and \cite{ouedraogo2023feasibility} have pointed out difficulties related to the high complexity of optimization problems, especially with increased variables and constraints. These complexities often result in a longer computation time, which is not feasible for real-time operation and day-ahead scheduling. Studies in \cite{reza2023uncertainty} and \cite{kang2023bi} highlighted the inherent uncertainties of renewable energy sources and load demand, which have significant impacts on the optimization results. Such uncertainties often lead to sub-optimal or even infeasible solutions in practical operations. Furthermore, \cite{nawaz2023distributed} and \cite{reniers2023digital} show the challenges in considering detailed battery characteristics, such as battery degradation and state of charge limits, in the optimization process. These detailed features could significantly affect the battery lifetime and the overall system performance but are often neglected due to the increased complexity they bring to the optimization problem. 

While numerous studies have tackled various facets of BESS control, an oft-neglected component is the dynamic pricing of tariffs. A significant portion of existing literature assumes a static pricing model \cite{serrano2017fixed} and \cite{boomsma2012renewable}, with tariffs remaining fixed throughout the day. Such an assumption can lead to gaps in the realistic applicability of their proposed solutions. Recognizing this oversight, our research places a specific emphasis on considering the dynamic nature of tariff pricing. We delve into its oscillations and evaluate the subsequent impact on BESS control performance across a 24-hour cycle deployed within smart buildings.

Our research aims to address several critical gaps in the existing approaches to BESS scheduling. Primarily, we focus on reducing the computational burden associated with training controllers to optimize battery and grid set points amid uncertainties in generation, demand, and tariff profiles. Another significant gap is ensuring that soft constraints, such as SOC thresholds, are consistently respected to prevent operational inefficiencies and potential batteries damage \cite{chen2022reinforcement}. Furthermore, our approach seeks to reduce electrical costs compared to conventional optimizers by improving the economic efficiency of energy storage usage. Lastly, we aim to enhance the reliability of model-free algorithms, ensuring their efficacy in real-time BESS operations. These improvements are expected to provide a more robust, cost-effective, and reliable framework for managing BESS in dynamic and uncertain energy markets.

In light of these limitations, our paper embarks on a two-fold approach. Initially, we employ a gradient-based method using Adam optimizer to verify feasible solutions for BESS control. Following this, we introduce a DRL framework, which offers the potential to not only alleviate the computational burden but also enhance our ability to discover superior global solutions. By leveraging the advantages of these two approaches, we aim to bring significant improvements in the optimization process of BESS control. We specifically address uncertainty in variables like renewable energy generation, load demand and dynamic tariff pricing. We test our model's robustness against various uncertainty scenarios, demonstrating its ability to maintain reliable, efficient solutions, reinforcing its potential for smart buildings use. Additionally, our goal is to have SOC of over 50\% by day's end, utilizing a select day-ahead scheduling algorithm. This approach ensures operational robustness and minimizes battery health risks.


The main contributions of this work can be summarized as follows:

\begin{itemize}

    \item BESS control problem balances trade-offs between grid and battery set points in a non-convex framework using variable penalties at each timestep. Leveraging gradient-based optimization as a benchmark, the introduction of DRL enhances management amid high variability in solar PV, load, and tariffs. Our approach seeks global optimization, minimizes daily electricity bills, and improves preparedness for worst-case scenarios, demonstrating significant improvements over gradient-based optimization.
    
    \item Uncertainty in BESS input data is quantified with an entropy term introduced as a state variable, allowing the control agent to account for system uncertainties and ensure robust actions. 
    
    \item Enhancements to the DRL agent's learning process include logarithmic scaling of the reward function and a physics safety layer within the DRL algorithm. This safety layer ensures system integrity and enforces hard constraints such as power balance, enhancing the robustness and reliability of system control.

\end{itemize} 

The rest of the paper is structured as follows: Section II details the framework and data collection methodology. Section III presents the problem formulation. Section IV discusses the Adam optimizer-based gradient optimization and the proposed DRL approach. Section V delves into the simulation results of the benchmark method and the tested DRL approach. The final section wraps up with the study's main conclusions and outcomes.

\section{Framework and Data collection}

\subsection{Framework structure}
\begin{figure*}[h]
  \centering
 \includegraphics[width=1.0\linewidth]{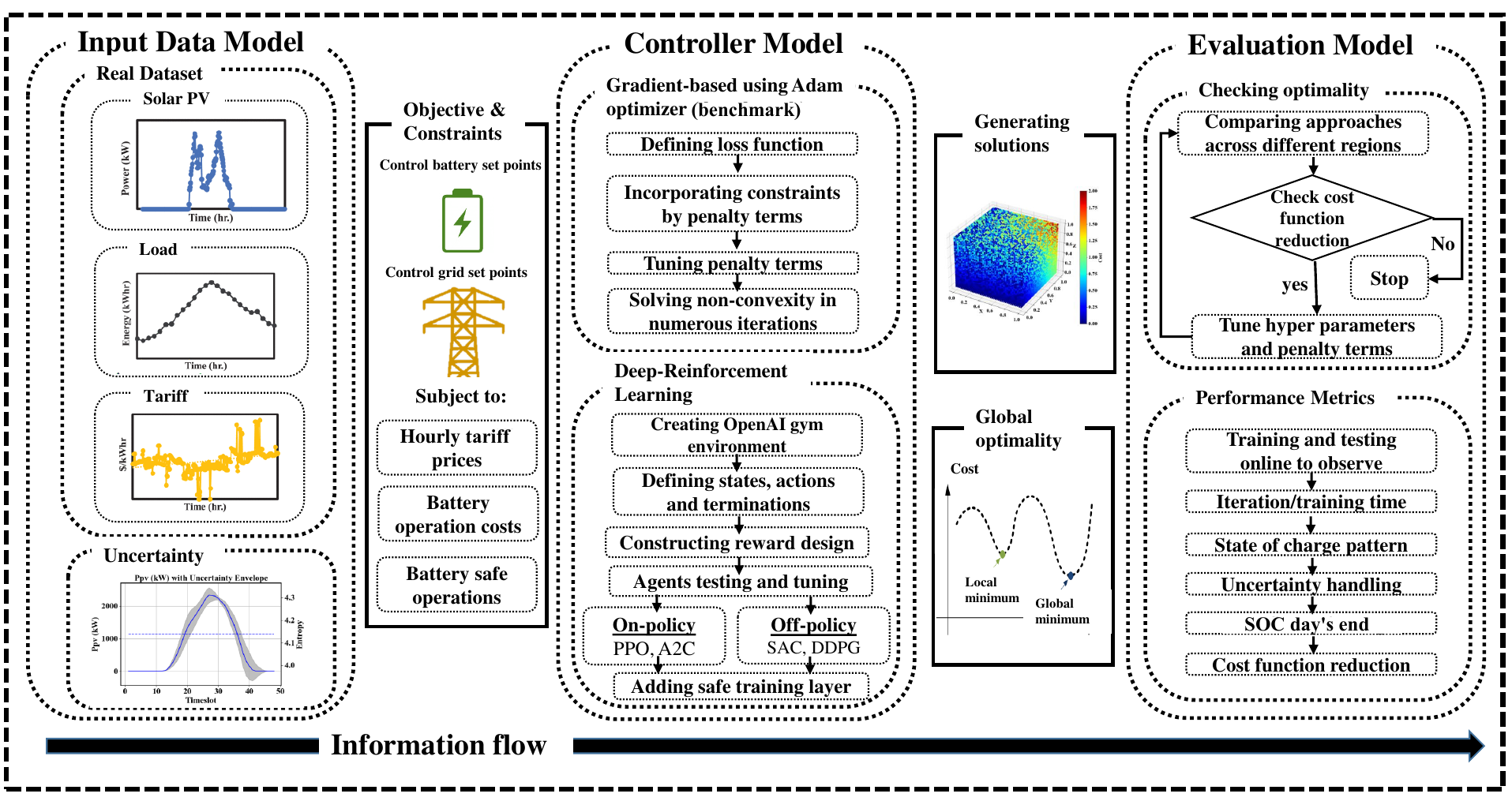}
 \centering
  \caption{Conceptual model for RMFEMF}
  \centering
 \label{Conceptual model for RMFEMF}
\end{figure*}

\textit{RMFEMF} (Robust Model-Free Energy Management Framework) is an innovative framework designed for the efficient and robust management of BESS within hybrid renewable energy settings in smart buildings as shown in Fig.\ref{Conceptual model for RMFEMF}. In contrast to other frameworks in the literature, as summarized in Table I, RMFEMF commences with high-resolution data collection from real field applications, i.e., five Australian sites, capturing energy loads, battery specs, PV generation, and grid tariffs to authentically simulate residential loads. It introduces a model-free simulation environment, which is adaptable and scalable, diverging from traditional model-based methods to dynamically represent BESS, PV systems, and grid interactions. Initial validation uses the Adam optimizer to test the solvability of optimal power exchanges in a residential setting. Following successful preliminary tests, RMFEMF employs the Soft Actor-Critic (SAC) algorithm \cite{SAC} to tackle the non-convex nature of the control problem and maintain battery SOC constraints. Finally, the framework undergoes calibration and experimental evaluation, adjusting hyperparameters and training off-line to ensure SOC remains above 50\% by the end of the day, thereby validating robustness without extensive experimental testing.

\begin{table*}[htbp] 
\caption{Comparison of Energy Management Frameworks in Literature}
\label{tab:comparison}
\centering
\begin{tabular}{@{}p{0.10\textwidth}p{0.08\textwidth}p{0.20\textwidth}p{0.12\textwidth}p{0.10\textwidth}p{0.10\textwidth}p{0.10\textwidth}p{0.08\textwidth}@{}} 
\toprule
\textbf{Reference} & \textbf{Problem Linearity} & \textbf{Objectives} & 
\textbf{Optimizer} & \textbf{Battery Control} & \textbf{Grid Control} & \textbf{Centralized} &
\textbf{Robust} \\
\midrule
\cite{nan2018optimal} & Linear & electricity cost &  N/A & $\times$ & $\times$ & $\times$ & $\times$ \\
\cite{nebuloni2023hierarchical} & Linear & battery replacement &  MILP, CPLEX & $\checkmark$ & $\times$ & $\times$ & $\times$ \\
\cite{javadi2020optimal} & Linear & electricity cost, user comfort &  MILP, CPLEX & $\times$ & $\times$ & $\times$ & $\times$ \\
\cite{hosseini2024optimal} & Non-linear & battery efficiency & fmincon & $\checkmark$ & $\checkmark$ & $\checkmark$ & $\times$ \\
\cite{revathi2024hybrid} & Non-linear & electricity cost, emissions, peak load & MAO, HBA & $\times$ & $\times$ & $\checkmark$ & $\times$ \\
\cite{hassaballah2024novel} & Non-linear & electricity price & fmincon & $\checkmark$ & $\checkmark$ & $\checkmark$ & $\times$ \\
\cite{hassan2022optimal} & Non-linear & electricity price & GA & $\times$ & $\times$ & $\checkmark$ & $\times$ \\
\cite{pinciroli2023optimal} & Non-linear & profit, battery maintenance & DRL-PPO & $\checkmark$ & $\times$ & $\checkmark$ & $\checkmark$ \\
\cite{huang2020deep} & Non-linear & profit & DRL-PPO & $\checkmark$ & $\times$ & $\checkmark$ & $\checkmark$ \\
\cite{nakabi2021deep} & Non-linear & profit & DRL-A3C & $\checkmark$ & $\checkmark$ & $\checkmark$ & $\checkmark$ \\
\cite{mohamed2023battery} & Non-linear & energy cost & DRL-CEM  & $\checkmark$ & $\times$ & $\checkmark$ & $\checkmark$ \\
\cite{guo2022real} & Non-linear & operation cost & DRL-PPO & $\checkmark$ & $\times$ & $\checkmark$ & $\checkmark$ \\
\cite{xiong2023meta} & Non-linear & total cost of customers & Meta-RL & $\checkmark$ & $\checkmark$ & $\times$ & $\checkmark$ \\
Proposed framework & Non-linear & energy cost, SOC day's end, computation time & DRL-SAC & $\checkmark$ & $\checkmark$ & $\checkmark$ & $\checkmark$ \\
\bottomrule
\end{tabular}
\end{table*}

\subsection{Datasets Collection}

In this paper, we leverage real-time datasets from five different locations in Australia to conduct a comprehensive analysis of the energy market. The utilization of these datasets provides valuable insights into the dynamics and patterns of energy consumption, solar PV profiles, and tariff prices in the selected regions. The demand dataset is obtained from the National Electricity Market Web (NEMWEB) platform \cite{demand_dataset}. This dataset enables the examination of electricity demand patterns, allowing for a deeper understanding of peak load periods and consumption trends. To capture the impact of solar power generation, we acquire the solar PV dataset from the NEMWEB Solar PV Database \cite{solar_pv_dataset}. This dataset provides valuable information on the solar PV profiles, facilitating the assessment of solar energy generation capacities and its integration into the electricity grid. In order to analyze the tariff prices and their influence on consumer behavior, we refer to the NEMWEB Tariff Price dataset \cite{energy_rrp_dataset}. This dataset allows for an examination of the dynamic pricing structures and their implications on energy consumption patterns. By focusing on five specific locations in Australia, namely New South Wales (NSW), Queensland (QLD), South Australia (SA), Tasmania (TAS), and Victoria (VIC), our study ensures a comprehensive representation of different states across Australia. The selected date of 13 June 2023, falling on a weekday, enabling an in-depth evaluation of solar PV profiles, Energy Region Reference Price (RRP), and load profile patterns in the studied regions. 



In our research, we primarily examine the Net System Load Profile (NSLP) \cite{aemo-load-profiles}, essential for understanding and analyzing electricity consumption patterns and distribution network dynamics. These profiles are calculated by MSATS (Metered Stand Alone Transport Solution) during every settlement aggregation run, then frozen weekly about 15 weeks post-settlement, and published sequentially on a designated webpage \cite{aemo-load-profiles}. This process includes determining the load shape based on the system profile of the distribution network where the basic meter is installed. By analyzing these profiles, we aim to enhance insights into load behaviors and assess the impact and optimization potential of controlled loads in the energy market.

\begin{figure}[h]
  \centering
  \includegraphics[width=1.05\columnwidth]{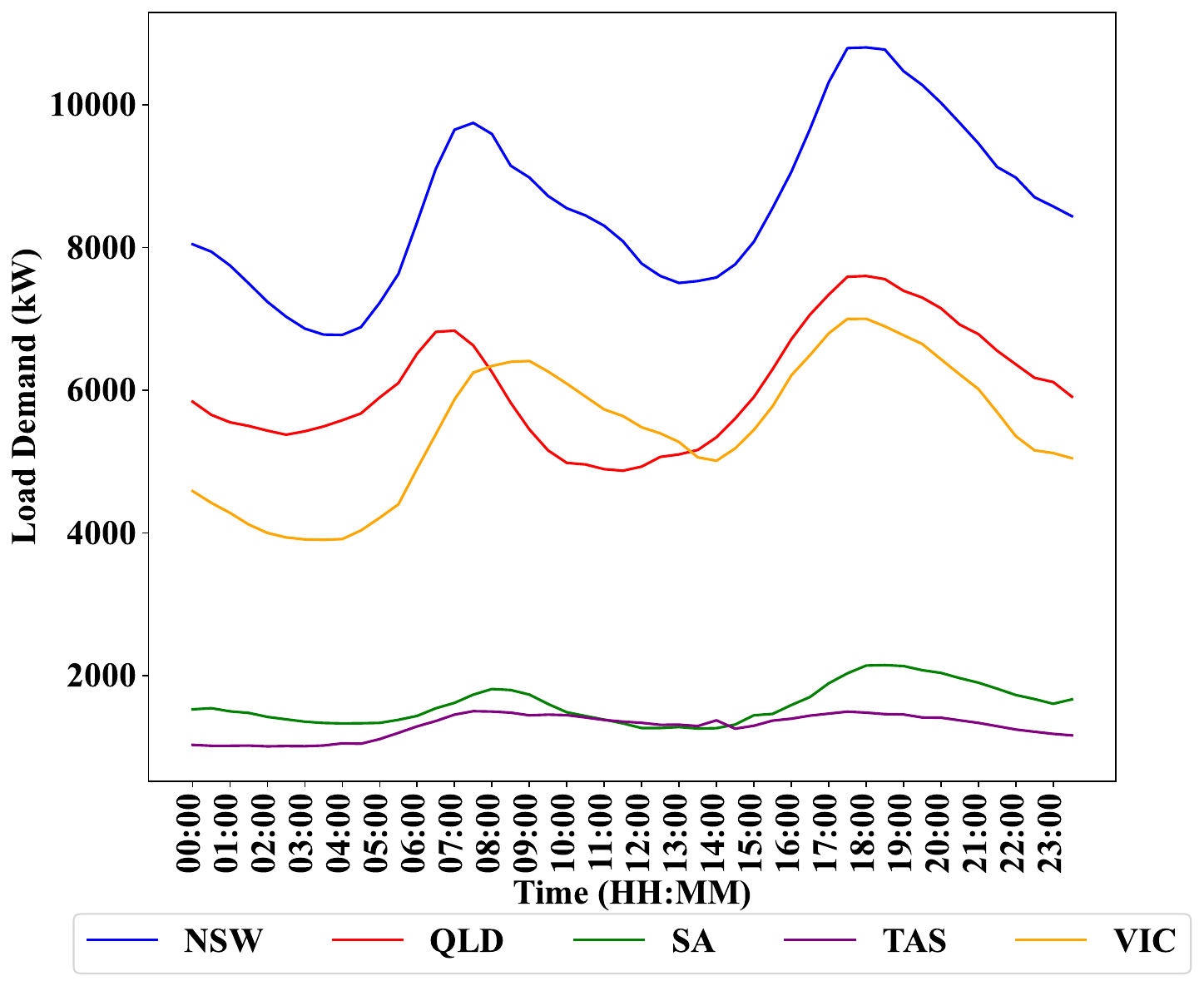}
  \centering
  \caption{Load Profiles for Australian States' districts}
  \centering
  \label{Load Profiles for Australian States' districts}
\end{figure}

\begin{figure}[h]
  \centering
  \includegraphics[width=1.05\columnwidth]{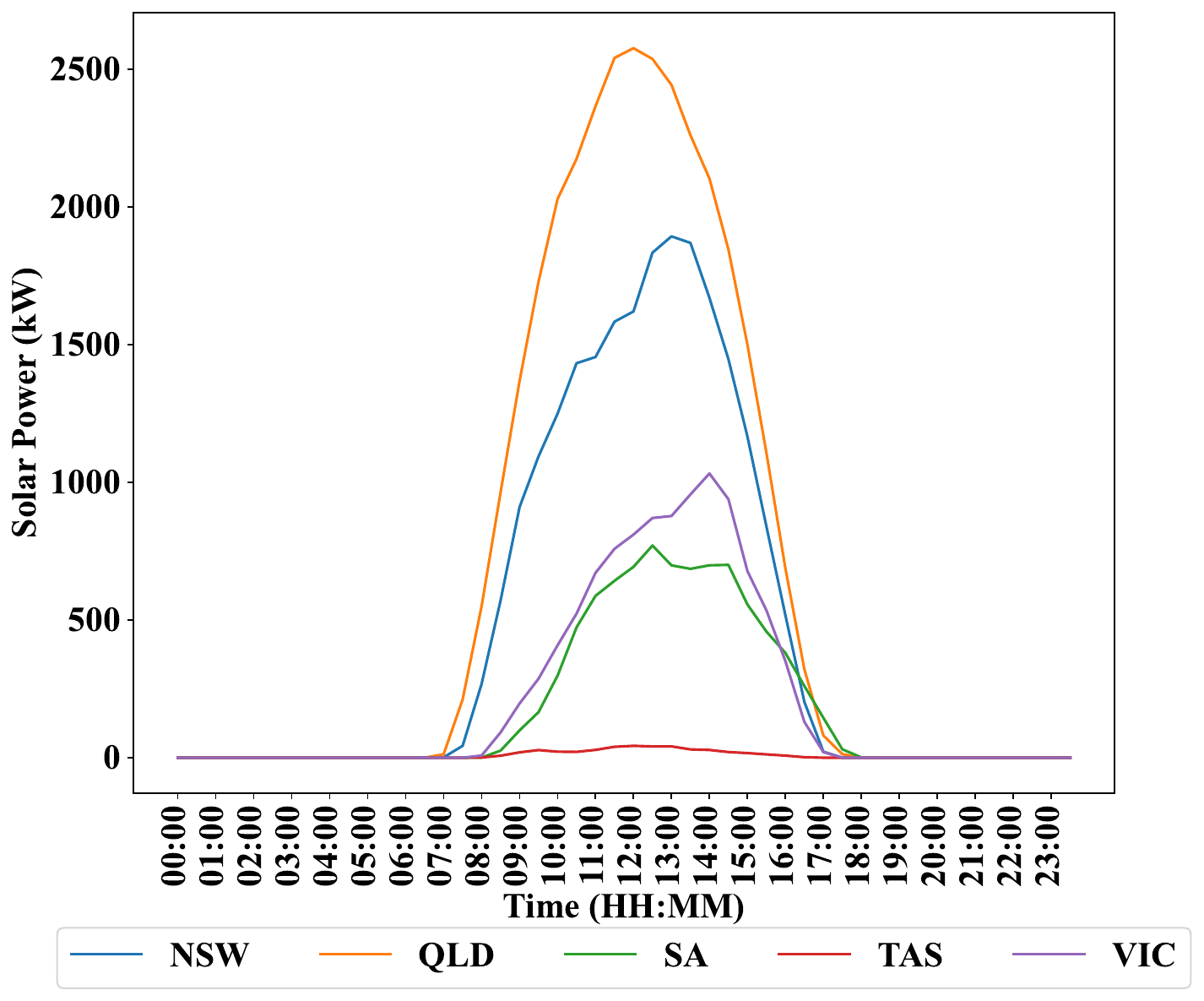}
  \centering
  \caption{Solar PV Profiles for Australian States' districts}
  \centering
  \label{Solar PV Profiles for Australian States}
\end{figure}

\begin{figure}[h]
  \centering
  \includegraphics[width=1.05\columnwidth]{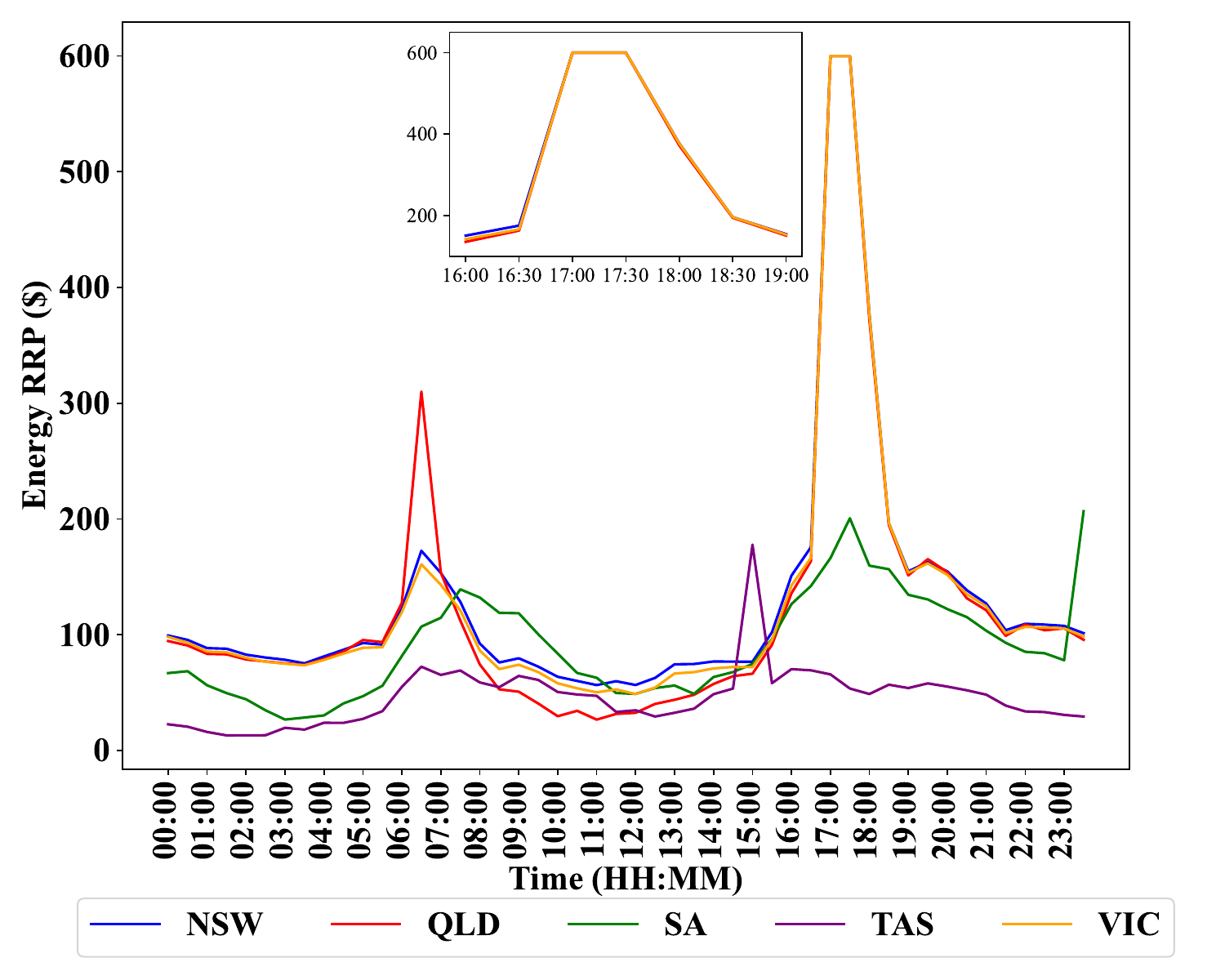}
  \centering
  \caption{Energy Region Reference Price (RRP) Profiles for Australian States' districts}
  \centering
  \label{Energy RRP Profiles for Australian States}
\end{figure}

It is worth mentioning that due to the inherent capacity constraints of BESS, we adapt the aggregate residential load profiles and solar PV profiles provided by NEM which indicate energy capacities for each state in the MW range. To make these macro-level data compatible with our micro-level study and after recognizing that behind-the-meter data are often inaccessible due to privacy concerns, we employ a scale-down approach that allows us to utilize these patterns in the context of a single residential district as displayed in Fig. \ref{Load Profiles for Australian States' districts} to Fig. \ref{Energy RRP Profiles for Australian States}. By doing so, we are able to conduct our BESS control simulations using realistically informed, albeit scaled-down, energy profiles.

\section{Problem Formulation}

The problem is formulated based on deducing a scheduled operation for BESS and grid supply to obtain a highly optimized performance for the energy dispatch and maintain the added economical value of BESS. This formulation extends the models proposed in \cite{selim2022optimal} and \cite{selim2023optimal} by introducing penalty terms as variable decision factors, rather than constants as assumed in previous studies. The assumption of variable penalty terms, enables the model to adjust the trade-off relationship automatically for each timestep, based on the operational conditions of the system. This approach ensures that the trade-offs are not rigidly fixed, as they would be with constant weights, but are instead dynamically tuned to reflect real-time system requirements.

\begin{equation}
\begin{aligned}
& \underset{\alpha^g_t, \alpha^b_t, P^g_t, P^b_t}{\text{minimize}}
& & \sum_{t=0}^T \alpha^g_t \cdot C_t^g \cdot P_t^g + \sum_{t=0}^T \alpha^b_t \cdot C_t^b \cdot P_t^b, \quad \forall t \in T
\end{aligned}
\end{equation}

subject to:
\begin{equation}  
P_t^{g}+ P_t^{b}+P_t^{pv}=P_t^d+P_t^{unc}, \quad \forall t \in T
\end{equation}
\begin{equation}  
SOC_{t+1} = SOC_t + \Delta T \left(\frac{P_t^b}{E_t^b}\right), \quad \forall t \in T
\end{equation}
\begin{equation}  
0 < SOC_t < 1, \quad \forall t \in T
\end{equation}
\begin{equation}  
\alpha^g_t + \alpha^b_t \leq \alpha^P, \quad \forall t \in T
\end{equation}
\begin{equation}  
\alpha^{g,\min}_t < \alpha^g_t < \alpha^{g,\max}_t, \quad \forall t \in T
\end{equation}
\begin{equation}  
\alpha^{b,\min}_t < \alpha^b_t < \alpha^{b,\max}_t, \quad \forall t \in T
\end{equation}
\begin{equation}  
-\left|P_t^{b, \min }\right| < P_t^b < \left|P_t^{b, \max }\right|, \quad \forall t \in T
\end{equation}
\begin{equation}  
-\left|P_t^{g, \min }\right| < P_t^g < \left|P_t^{g, \max }\right|, \quad \forall t \in T.
\end{equation}

The objective function shown in (1) aims to minimize the grid power imported (i.e., minimizing tariff prices) and also minimize the rate of energy dispatched by the battery for a long life cycle. Action variables for the studied horizon of the system are identified as follows:   ${P}_{t}^{b}$ is the power dispatched by the energy storage system within the storage limits of $P_{t}^{ {b,max }}$ and $P_{t}^{ {b,min}}$.  ${P}_{t}^{g}$ is the power supplied by the grid within the $P_{t}^{{g,max }}$ and $P_{t}^{{g,min }}$ limits.  State variables here include $C_{t}^{g}$ and $C_{t}^{b}$, which  represent the immediate tariff price for electricity and the operational cost of the battery, respectively. ${P}_{t}^{d}$ is the total power demand for the system, ${P}_{t}^{pv}$ is the rooftop solar power and ${P}_{t}^{unc.}$ is the power needed to balance uncertainties in generation and demand.  The newly introduced action variables, $\alpha_{g}$ and $\alpha_{b}$, are of significant importance as they serve as variable penalty terms. They affect the percentage of power-sharing between the grid and batteries, respectively, thereby playing a crucial role in optimizing the system's performance. These terms are variable in time and require precise control for optimized performance. The time-variability and the need to control of these terms add complexity to the problem, making it non-convex. Some studies in \cite{arghandeh2014economic} and \cite{selim2022optimal} assume these penalty terms to be constant for the day-ahead scheduling problem. However, in this paper, we investigate the control of these terms to understand their impact on the scheduling and optimization of BESS operation and management. This investigation aims to address the challenges posed by the non-convex nature of the problem.

\subsection{Uncertainty Types}


Uncertainties in energy systems like power demand, solar power generation, and grid tariffs are modeled using distributions such as normal, uniform, exponential, log-normal, and Beta, due to varying policies and market structures \cite{Ross2014, meeus2009}. These models assess the impacts on BESS control strategies with an uncertainty envelope of 10\% \cite{Hoppmann2014}. Variables in the model, denoted by \(x_i\), include \(P^{pv}_{t}\), \(P^d_t\), \(C^g_t\), and \(C^b_t\). The transformations applied to each variable are described through various probability distributions as follows:
\begin{equation}
u_{norm,i} = \mathcal{N}(\mu, \sigma) \cdot x_i, \quad \mathcal{N}(\mu, \sigma): \text{normal distribution}
\end{equation}
\begin{equation}
u_{uni,i} = \mathcal{U}(a, b) \cdot x_i, \quad \mathcal{U}(a, b): \text{uniform distribution}
\end{equation}
\begin{equation}
u_{exp,i} = (\mathcal{E}(\lambda) - \Delta) \cdot x_i, \quad \mathcal{E}(\lambda): \text{exponential distribution}
\end{equation}
\begin{equation}
u_{logn,i} = (\mathcal{LN}(\mu, \sigma) - \Delta) \cdot x_i, \quad \mathcal{LN}(\mu, \sigma): \text{log-normal distribution}
\end{equation}
\begin{equation}
u_{beta,i} = (\mathcal{B}(\alpha, \beta) - \Delta) \cdot \text{scale} \cdot x_i, \quad \mathcal{B}(\alpha, \beta): \text{beta distribution}
\end{equation}

In our model, variable transformations are applied through several probability distributions. The variable \(u_{norm,i}\) is transformed according to the normal distribution \(\mathcal{N}(\mu, \sigma)\), with specified mean \(\mu\) and standard deviation \(\sigma\). For uniform effects, \(u_{uni,i}\) utilizes the uniform distribution \(\mathcal{U}(a, b)\), spanning from lower bound \(a\) to upper bound \(b\). The exponential distribution effects are captured by \(u_{exp,i}\), following \(\mathcal{E}(\lambda)\) adjusted by a constant \(\Delta\). Log-normal distribution transformations are applied to \(u_{logn,i}\), modeled by \(\mathcal{LN}(\mu, \sigma)\) and similarly adjusted. Lastly, \(u_{beta,i}\) reflects beta distribution effects through \(\mathcal{B}(\alpha, \beta)\), where it is both scaled and shifted by \(\Delta\) to achieve desired properties.

\subsection{Uncertainty Quantification}



\begin{figure}[!ht]
\centering
\includegraphics[width=1.0\columnwidth]{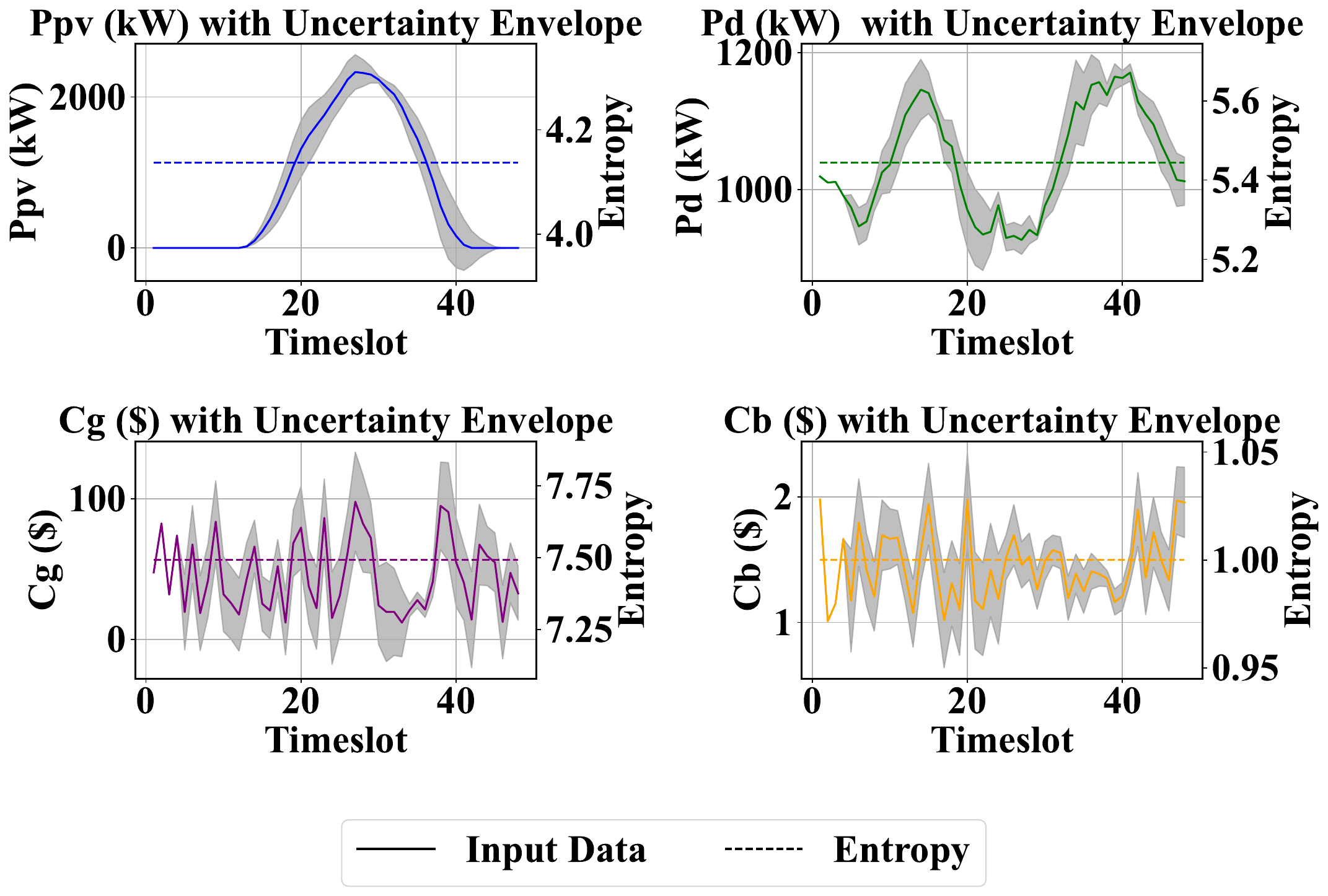}
\centering
\caption{Uncertainty envelope and entropy computed for the input data\label{Uncertainty envelope}}
\end{figure}

Entropy quantifies the uncertainty in a probability distribution \cite{entropy}. It is valuable for comparing probability distributions to assess uncertainty levels. Calculating entropy begins by defining the variable's probability distribution, such as a Gaussian,  discretizing this distribution into bins by segmenting the probability density function and computing each bin's probability through integration over its segment. Entropy is then calculated with the formula \cite{shannon1948mathematical}:
\begin{equation}
\begin{gathered}
\textit{$H(X)$ = $-\sum_{i=1}^{n} p_i \log_2(p_i)$,}
\end{gathered}
\end{equation}
where $H(X)$ represents the entropy of variable $X$, $p_i$ is the probability of the $i^{th}$ bin, and $n$ is the total number of bins. Uncertainty can be quantified using methods such as uniform distribution or entropy index, as illustrated in Fig. \ref{Uncertainty envelope}. These approaches offer intuitive states for optimizers or agents in energy management systems, guiding control actions and enhancing system robustness against uncertainties.

\section{Proposed control approaches}
\subsubsection{\textbf{Benchmark approach}}
Gradient-based optimization, despite its limitations such as susceptibility to local minima, remains a popular method for tackling non-linear, non-convex optimization challenges by iteratively updating model parameters to minimize the objective function. Strategies to mitigate getting stuck in local minima include varied initialization, noise addition, and advanced algorithms capable of escaping such minima. In this context, the BESS problem is formulated with decision variables, where the complex, non-convex optimization landscape with numerous local optima is evident. The Adam optimizer \cite{Adam} is chosen for its effectiveness in managing noisy and non-smooth data, with relatively easy-to-tune hyperparameters. In our implementation, the optimization algorithm (Algorithm 1) in \nameref{sec:appendix} includes variables $P^g_t$, $P^b_t$, and penalty multipliers $\alpha^g_t$, $\alpha^b_t$. The loss function $loss_{\text{fn}}$, variable $var$, constraints $bounds$, initial learning rate $init_{\text{lr}}$, decay steps $d_{\text{steps}}$, decay rate $d_{\text{rate}}$, and number of epochs $n_{\text{epochs}}$ are defined. Optimization is performed iteratively using an Adam optimizer with a learning rate scheduler.

\subsection{Deep Reinforcement Learning Approach}
\subsubsection{\textbf{Proposed approach}}
To complete the RMFEMF framework, after identifying potential solutions through the gradient-based methods, we recognize the necessity for further refinement and exploration of optimization techniques. Consequently, we turn to employ the DRL algorithm in our framework. The use of DRL provides an opportunity to harness its capabilities in dealing with complex decision-making scenarios and potentially offers a more adaptive and efficient learning approach. This incorporation aims to investigate if DRL can achieve a more harmonized balance in minimizing costs, ensuring no SOC violations, and maintaining the end-of-day SOC at acceptable levels, particularly in light of the observed shortcomings of the gradient-based methods (shown later in the results section) in achieving the desired end-of-period SOC. Through the integration of DRL, we aspire to enhance the robustness and performance of the RMFEMF framework and contribute towards more informed and efficient energy management strategies for the regions under study. 

To learn the optimal policy, we adopt SAC, a state-of-the-art DRL algorithm \cite{SAC} that combines the strengths of off-policy learning and entropy maximization as shown in Fig. 6. The SAC algorithm enables the model to learn the stochastic policy that is optimal in terms of the expected return and entropy, a measure of randomness in the policy. This algorithm is particularly suitable for our problem due to its ability to handle continuous action spaces and its capacity to balance exploration and exploitation, a critical aspect in the presence of non-convexities and multiple local optima. SAC algorithm is particularly suited for the optimal scheduling of BESS in an energy grid. Here, we explore how the components of SAC can be specifically tailored to address this application:

(a) \textbf{Markov Decision Process (MDP) Framework and Objective for BESS:}
In the context of BESS, the states (\(s_t\)) represent the current charge levels, time of day, electricity prices, and demand forecasts. Actions (\(a_t\)) would be the amount of energy to charge or discharge. The objective of Soft Actor-Critic (SAC) in this scenario includes maximizing financial returns by smartly trading energy while ensuring battery health by minimizing unnecessary cycling:
\begin{equation}
J(\pi) = \mathbb{E}_{(s_t, a_t) \sim \rho_\pi} \left[ \sum_{t=0}^{\infty} \gamma^t \left( \text{Profit}(s_t, a_t) + \alpha \mathcal{H}(\pi(\cdot | s_t)) \right) \right]
\end{equation}
where \(J(\pi)\) is the objective function, \(\pi\) is the policy, \(\rho_\pi\) is the state-action visitation distribution, \(\gamma\) is the discount factor, \(\alpha\) is the temperature parameter that determines the stochasticity of the policy, and \(\mathcal{H}(\pi(\cdot | s_t))\) is the entropy of the policy. This formulation encourages the exploration of various charging and discharging strategies to identify the most optimal operational patterns.

(b) \textbf{Policy Network for BESS Decision-Making:}
The policy network (\(\pi_\theta\)) dictates how the BESS decides between charging, discharging, or maintaining its current state based on the input state (\(s_t\)):
\begin{equation}
\pi_\theta(a_t | s_t) \propto \exp\left(\frac{1}{\alpha}(Q(s_t, a_t) - \log \pi_\theta(a_t | s_t))\right)
\end{equation}
where \(\pi_\theta\) is the policy parameterized by \(\theta\), \(a_t\) is the action at time \(t\), \(Q(s_t, a_t)\) is the action-value function. This stochastic policy helps in exploring a diverse set of actions, which is crucial for adapting to the highly variable energy prices and demand patterns.

(c) \textbf{Critic Networks for BESS Value Estimation:}
The twin critic networks estimate the future value of current actions, helping to balance immediate rewards (like peak shaving benefits) with long-term outcomes (such as battery health and degradation costs):
\begin{equation}
\begin{split}
    Q(s_t, a_t) = & \text{Immediate Reward}(s_t, a_t) \\
    & + \gamma \mathbb{E}_{s_{t+1} \sim p} \left[ \min_{i=1,2} Q_{\text{target}, i}(s_{t+1}, \pi(s_{t+1})) \right]
\end{split}
\end{equation}
where \(Q(s_t, a_t)\) is the action-value function, \(\gamma\) is the discount factor, \(s_{t+1}\) is the next state, \(p\) is the state transition probability, and \(Q_{\text{target}, i}\) are the target critic networks. This setup is ideal for evaluating the trade-offs between different operational strategies, including the timing of charging and discharging actions to maximize utility rates or to participate in demand response programs.

(d) \textbf{Experience Replay and Mini-Batch Learning in BESS Management:}
The use of a replay buffer allows the SAC to learn from a wide range of historical scenarios, which is invaluable given the fluctuating nature of energy markets and grid demands:
\begin{equation}
\text{Mini-batch loss} = \frac{1}{|\mathcal{B}|} \sum_{(s_t, a_t, r_t, s_{t+1}, d_t) \in \mathcal{B}} \left[ Q(s_t, a_t) - y_t \right]^2
\end{equation}
where \(\mathcal{B}\) is the mini-batch sampled from the replay buffer, \(r_t\) is the reward, \(d_t\) is the done signal, and \(y_t\) is the target value. Training from these mini-batches helps the system refine its strategy over time, ensuring robustness and adaptability to new information and changing grid conditions.

(e) \textbf{Entropy-Based Exploration for Optimal BESS Operation:}
    By maximizing entropy, the SAC ensures that the BESS does not settle too quickly on a potentially suboptimal charging/discharging schedule, especially in the face of uncertain future energy prices and load demands.

\begin{figure}[!ht]
\centering
\includegraphics[width=1.04\columnwidth]{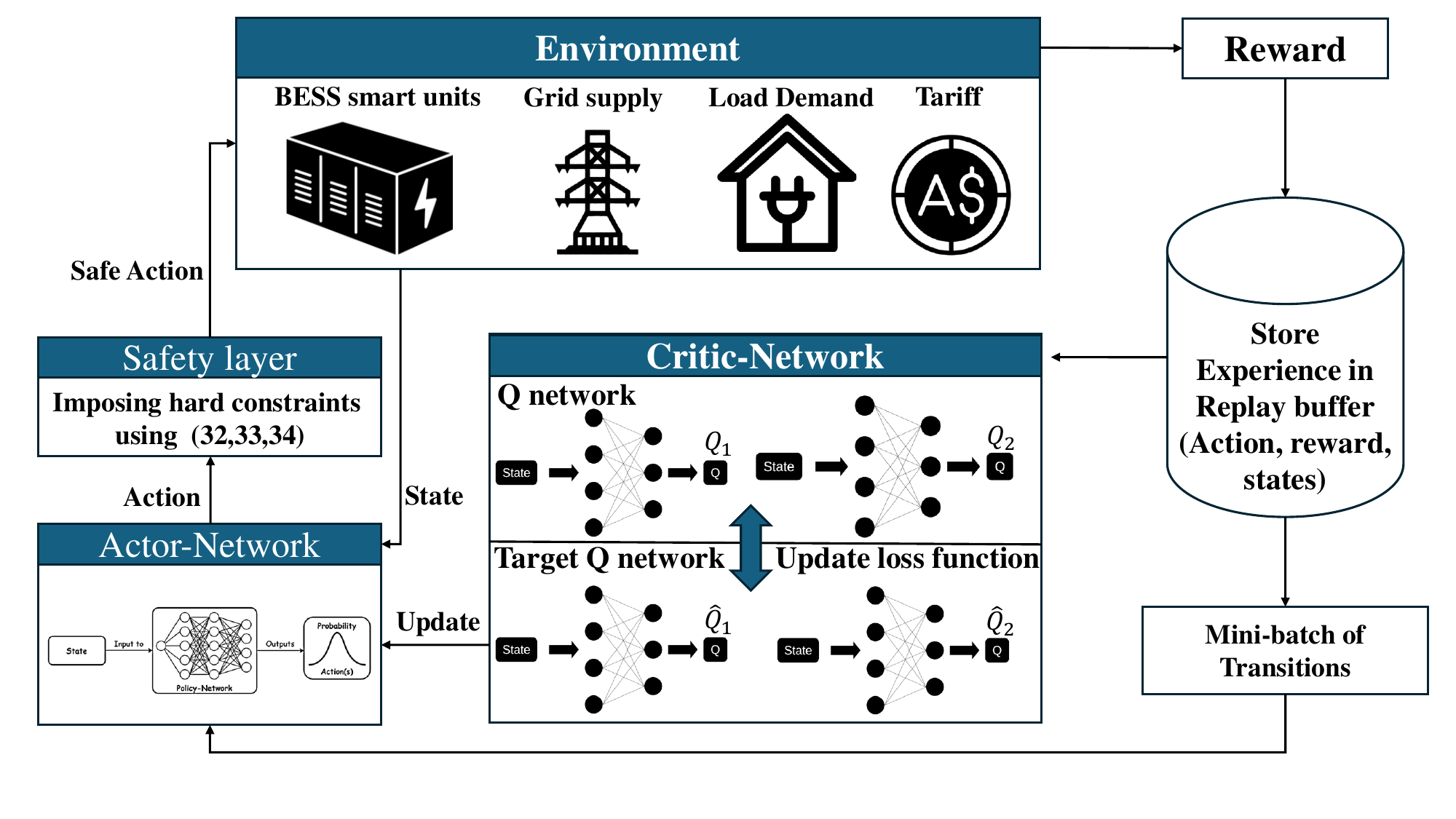}
\centering
\caption{Proposed SAC algorithm to control BESS }
\end{figure}



\subsubsection{\textbf{Environment}}
The BESS control environment is formulated as a MDP as shown in Fig. 7. During each time step, the agent decides either the grid power or battery power to be utilized, and it receives a reward that is inversely proportional to the total cost, including penalties for any constraint violations. Once the agent has learned to perform actions that maximize its cumulative reward, it effectively learns the optimal strategy for grid power and battery power utilization. The model is trained over a significant number of time steps, allowing it to interact with the environment and refine its policy over time. 
We define a custom OpenAI gym environment \cite{brockman2016openai}.

\begin{figure}[!ht]
\centering
\includegraphics[width=1.0\columnwidth]{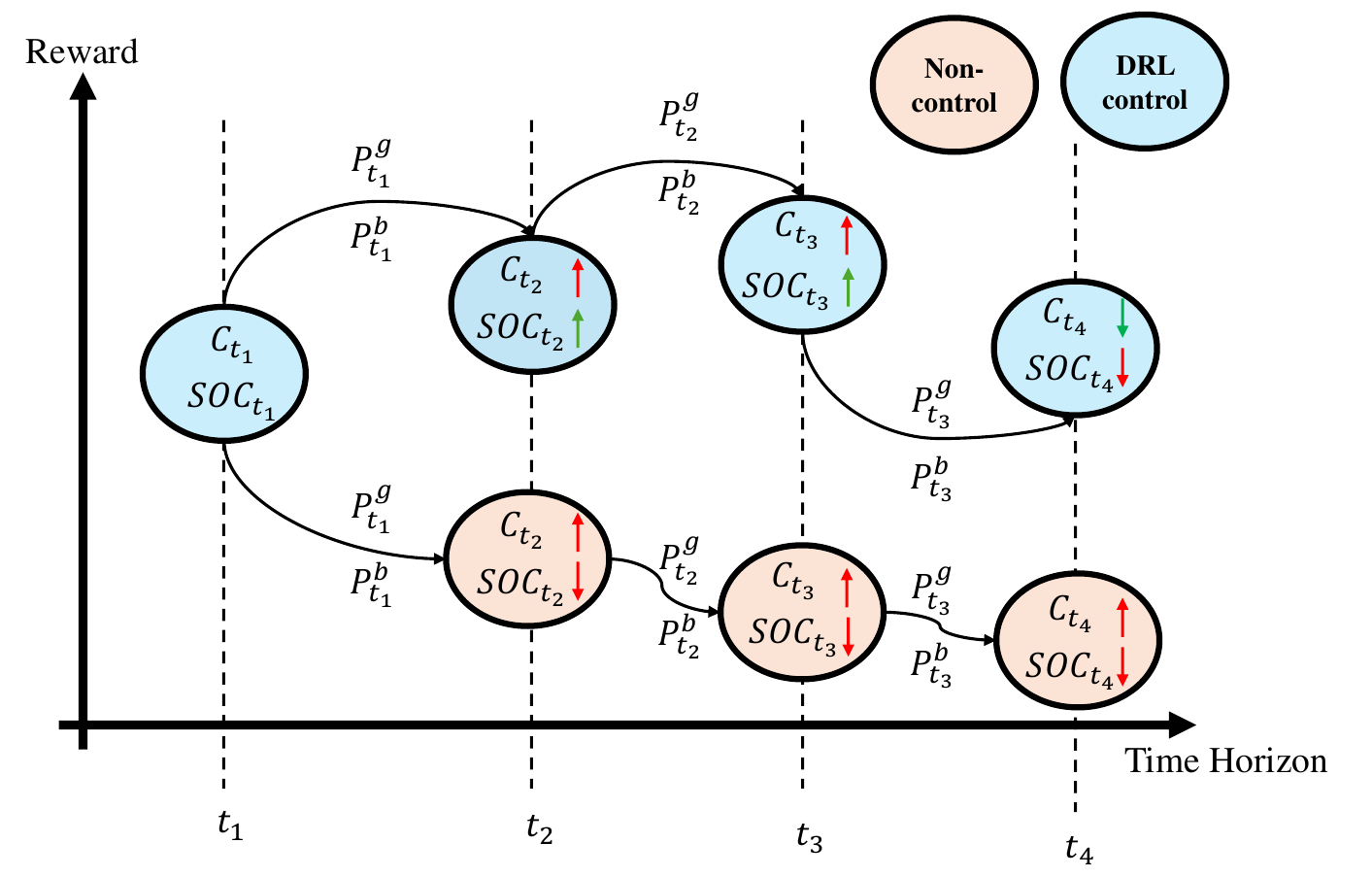}
\centering
\caption{MDP during BESS control}
\end{figure}

Our approach towards designing an energy management framework utilizes  the stable-baselines3 library \cite{raffin2021stable}, which includes algorithms such as Proximal Policy Optimization (PPO) \cite{PPO}, Advantage Actor-Critic (A2C) \cite{A2C}, Deep Deterministic Policy Gradient (DDPG) \cite{DDPG}, and Twin Delayed DDPG (TD3) \cite{TD3}.  This environmental state includes photovoltaic power generation, power demand, grid and battery costs, power uncertainty, and SOC of the battery: 

\begin{equation}
s_t = [{P}_{t}^{pv}, {P}_{t}^{d}, {C}_{t}^{g}, {C}_{t}^{b}, {P}_{t}^{unc}, {SOC}_{t}],
\end{equation}
or 
\begin{equation}
s_t = [{P}_{t}^{pv}, {P}_{t}^{d}, {C}_{t}^{g}, {C}_{t}^{b}, H(x), {SOC}_{t}].
\end{equation}

We keep the penalty terms fixed for $\alpha_{g}$ and $\alpha_b $ based on the grid search space, which also tunes them for the best possible value. This approach can significantly improve the training phase of the DRL algorithm. 

\begin{equation}
a_t = [{P}_{t}^{g}, {P}_{t}^{b}].
\end{equation}


\subsubsection{\textbf{Cost Function }}

Our agent is trained by optimizing a reward function, which penalizes power imbalance, deviation of SOC from its limits and a logarithmic function of the total cost. This cost function is based on the main objective function defined in (1) and will be subject to some modifications in the reward function designed later. In our exploration of the parameter space for the $alpha$ values, we discover that setting the penalty terms to a consistent value of 100 can substantially improve the results as shown in (19). While we continue to adhere to the principle of controlling penalty terms over time, approximating them to constant values throughout the entire horizon can remarkably enhance the training phase of the DRL. In that way, the DRL algorithm can efficiently focus on minimizing deviations from the target power set points and impose a fixed penalty of 100 for any discrepancies. This streamlines the optimization process, stabilizes the learning, and facilitates better control over power set points. The choice of 100 is empirically grounded in the analysis of grid search results, which indicates improved performance and convergence with constant penalties.
\begin{equation}
C_t = \alpha^g_t \times {P}_{t}^{g} \times {C}_{t}^{g} + \alpha^b_t \times {P}_{t}^{b} \times {C}_{t}^{b}.
\end{equation}

The power balance is calculated as the  difference between total power demand and total power supply as follows:

\begin{equation}
{P}_{t}^{bl} = ({P}_{t}^{d} + {P}_{t}^{unc.}) - ({P}_{t}^{g} + {P}_{t}^{b} + {P}_{t}^{pv}).
\end{equation}

\subsubsection{\textbf{Normalization of Input Parameters}}
In order to ensure that the input parameters are in a consistent range, we perform a normalization step by dividing each input parameter by a fixed constant value. This normalization process scales the input parameters to a uniform scale. We normalize these parameters to obtain the normalized values ${P}_{t}^{gn}$, ${P}_{t}^{bn}$, ${P}_{t}^{pvn}$, ${P}_{t}^{dn}$ and ${P}_{t}^{bln}$ by dividing each input parameter by a fixed constant value. This can be expressed using the following equation:

\begin{equation}
{X}_{\text{norm}} = \frac{{X}}{{k}},
\end{equation}

\noindent where ${X}_{\text{norm}}$ represents the normalized value of the input parameter ${X}$, and $k$ is the fixed constant used for normalization.  This normalization step helps in achieving better convergence during the training process.

\subsubsection{\textbf{Reward Design}}

The reward function in consideration is given by:
\begin{equation}
    r_t = -\log(C_t + 10^{-3}) - \alpha_{s} \times S_p - \alpha_{l} \times S_{lh} - \alpha_{p} \times P^{bl}_{t},
\end{equation}

    In the proposed model, \(r_t\) represents the reward at time \(t\), and \(C_t\) denotes the cost at the same time point. The model also incorporates several weighting factors for different penalty terms: \(\alpha_{s}\) is the weight assigned to SOC penalty term, which varies between 1 and 2 in different runs; \(\alpha_{l}\) represents the weight for the SOC lower half penalty term, also ranging between 1 and 2; and \(\alpha_{p}\) is the weight for the power balance penalty term, varying significantly between 100 and 1000 depending on each run. The other terms are computed as follows: 
\begin{itemize}
   \item \(S_p\) is the SOC penalty term, defined as:
  \begin{align}
        S_p = \eta \times \bigg(&\max(0, SOC_{\text{min}} - SOC_{t}) \nonumber \\
        &+ \max(0, SOC_{t} - SOC_{\text{max}})\bigg),
\end{align}

    where \(\eta\) is the penalty weight for the SOC term, \(SOC_{\text{min}}\) is the minimum SOC threshold, \(SOC_{\text{max}}\) is the maximum SOC threshold, and \(SOC_{t}\) is the SOC value at time \(t\);

\item \(S_{lh}\) is the SOC lower half penalty term which ensures that the battery has at least 50\% SOC reserve  for the next day, and having the same weight term of $\eta$ as following:
    \begin{equation}
        S_{lh} = \eta \times \max(0, 0.5 - SOC_{t});
    \end{equation}

\item \(P_{b}\) is the power balance penalty term, defined as:
    \begin{equation}
        P_{b} = \beta \times P^{bl}_{t},
    \end{equation}
    where \(\beta\) is the penalty weight for power balance.

\end{itemize}

The use of the logarithm function in the reward function can be attributed to the following reasons: a) \textbf{Smoothing and Scaling:} The logarithm function grows slowly as its argument increases, which is beneficial for compressing the scale of rewards, especially when \( C_t \) can take large values, thus avoiding very high magnitudes in the rewards that might affect the stability of training in reinforcement learning algorithms. b) \textbf{Encouraging Improvement:} Additionally, the logarithm function is concave and emphasizes relative improvements when the values are small, making large values less sensitive. In this scenario, for small and positive costs \( C_t \), the negative logarithm would result in substantially negative rewards for minor cost improvements. c) \textbf{Avoiding Division by Zero:} Furthermore, to prevent undefined operations, the term \( 10^{-3} \) is added within the logarithm to avoid taking the logarithm of zero. This small positive value ensures that the logarithm remains well-defined even if \( C_t \) is zero.

The goal of the agent is to find a policy that minimizes the total cost, while maintaining SOC within its limits and keeping the power balance as close to zero as possible. Table \ref{tab:my_label} and Table \ref{table:SAC_summary} show the structure of the proposed SAC-DRL agent and its bounded constraints.

\begin{table}[h!]
    \centering
    \caption{Hard and soft constraints in the DRL environment}
    \begin{tabular}{ccc}
        \hline
        Constraint & Hard & Soft \\
        \hline
        Action
        space upper and lower boundaries & $\checkmark$ & \\
        State space upper and lower boundaries & $\checkmark$ & \\
        Power balance & $\checkmark$ & \\
        SOC limit & & $\checkmark$ \\
        SOC day's end & & $\checkmark$ \\
        \hline
    \end{tabular}
    \label{tab:my_label}
\end{table}

\begin{table}[ht]
    \centering
    \caption{Key aspects of the SAC algorithm used in the code.}
    \label{table:SAC_summary}
    \begin{tabular}{cc}
        \toprule
        \textbf{Attribute} & \textbf{Value} \\
        \midrule
        Algorithm name & Soft Actor-Critic (SAC) \\
        Policy & Stochastic \\
        Entropy regularization & Yes \\
        Action space & Continuous \\
        Network architecture & 2 hidden layers, 64 units each \\
        Learning rate & 0.0005 \\
        Buffer size & 10000 \\
        Learning starts & 500 \\
        \bottomrule
    \end{tabular}
\end{table}

\subsubsection{\textbf{Algorithmic Enforcement of Power Balance}}

\begin{figure}[!ht]
\centering
\includegraphics[width=0.75\columnwidth]{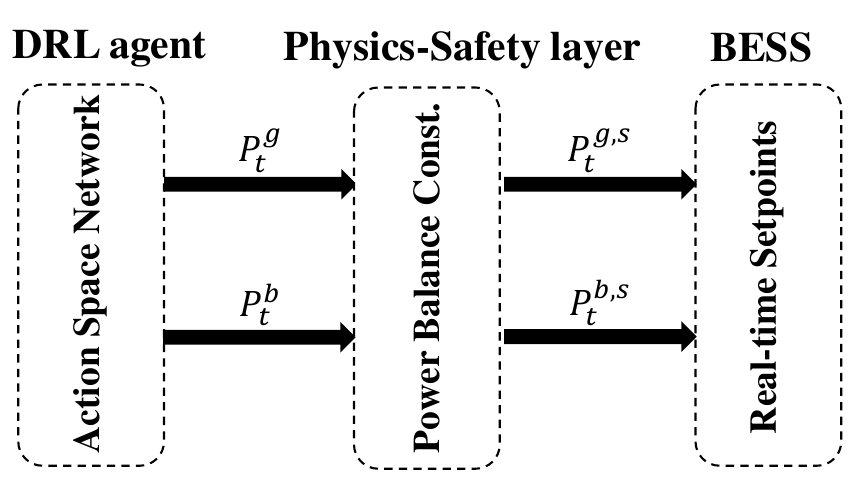}
\centering
\caption{Physics-safety layer\label{Safety Constain Idea}}
\end{figure}

Due to the model-free nature of DRL, it is observed that some hard constraints are violated during simulations. This is attributed to the fact that DRL algorithms learn from interactions without necessarily considering the underlying physical model or constraints. To address this issue, a physics-based safety layer is introduced into the system as shown in Fig.\ref{Safety Constain Idea}. This safety layer performs instantaneous action corrections to ensure that the real-time set points executed by the BESS comply with the physical constraints. This integration enhances the reliability and safety of the control actions, especially in a real-world deployment where the constraint adherence is crucial. To enforce the power balance constraint, we dynamically adjust the power drawn from the DRL agent as follows.

\begin{enumerate}[label=(\roman*)]

    \item Calculate total power generation, $T_p$, including photovoltaic power:
    \begin{equation}
        T_p = {P}_{t}^{g} + {P}_{t}^{b} + {P}_{t}^{pv}.
    \end{equation}
    
    \item Check if there is an imbalance between the total power generation, $T_p$, and the adjusted power demand, $A_d$. If $T_p\neq A_d$, proceed to step iii.
    
    \item Calculate the power, $R_p$, required from both batteries and grid to achieve balance:
    \begin{equation}
        R_p = A_d - {P}_{t}^{pv}.
    \end{equation}
    
    \item Calculate a scaling factor, $\kappa$:
    \begin{equation}
        \kappa = \frac{R_p}{{P}_{t}^{g} + {P}_{t}^{b}}     .
    \end{equation}
    
    \item Proportionally adjust the power drawn from the grid and the battery using the scaling factor, $\kappa$:
    \begin{align}
        {P}_{t}^{g,s} &= {P}_{t}^{g}\times \kappa, \\
        {P}_{t}^{b,s} &= {P}_{t}^{b} \times \kappa.
    \end{align}
\end{enumerate}

In the proposed system, the variables are defined as follows: \( T_p \) represents the total power generation, \( A_d = P_t^d + P_t^{unc} \) denotes the adjusted power demand at time \( t \), incorporating both deterministic (\( P_t^d \)) and uncertain (\( P_t^{unc} \)) components of power demand. \( R_p \) is the required power to achieve balance in the system, and \( \kappa \) is a scaling factor used to adjust the power outputs or inputs as needed to maintain system stability and efficiency. This algorithm ensures the enforcement of the power balance constraint by calculating and utilizing a scaling factor to adjust the power drawn from the grid and the battery based on the current power demand and generation.

\subsubsection{\textbf{Evaluation}}
Each agent is evaluated for its mean reward over several episodes to generate learning curves. The mean reward $\bar{R}$ for a given policy $\pi$ is estimated as:

\begin{equation}
\bar{R}(\pi) = \frac{1}{N} \sum_{i=1}^{N} R(\tau_i),
\end{equation}

\noindent where $N$ is the number of evaluation episodes, $\tau_i$ is the $i$-th episode, and $R(\tau_i)$ is the total reward of the $i$-th episode under policy $\pi$.

\section{Numerical Results}
\subsection{BESS Modeling}

Several  common control algorithms used in BESS systems include:
1) SOC control which ensures that the BESS operates within a desired SOC range by controlling the charging and discharging of the battery, and 2)
Power control which manages the power output of the BESS to meet the power demand of the system and ensure that the battery is not over-discharged or over-charged. In this paper, we implement these two practical methods by controlling the BESS with specific parameters, which are given in Table IV, to provide power set points, while respecting the system safety.
\begin{table}[h!]
\centering
\caption{Parameters used for training the optimizer.} 
\begin{tabular}{cc}
\toprule
\textbf{Parameter} & \textbf{Value} \\
\midrule
Number of epochs & 10,000,000 \\
Tolerance & $1 \times 10^{-5}$ \\
$P_{g,\text{min}}$ & 0 kW \\
$P_{g,\text{max}}$ & 5000 kW \\
$P_{b,\text{min}}$ & -1000 kW \\
$P_{b,\text{max}}$ & 1000 kW \\
Battery capacity & 20000 kWh \\
Initial SOC & 0.8 \\
Learning rate for Adam optimizer & 0.01 \\
Minimum and maximum values for $\alpha_g$ and $\alpha_b$  & 1, 1000 \\
\bottomrule
\end{tabular}
\label{table:parameters}
\end{table}

\subsection{Gradient-Based Optimization using Adam}

\begin{figure*}[p]
  \centering
  \raisebox{-0.9\totalheight}[0pt][0pt]{\rotatebox{90}{NSW}}
  \subfloat[]{\includegraphics[width=0.325\textwidth]{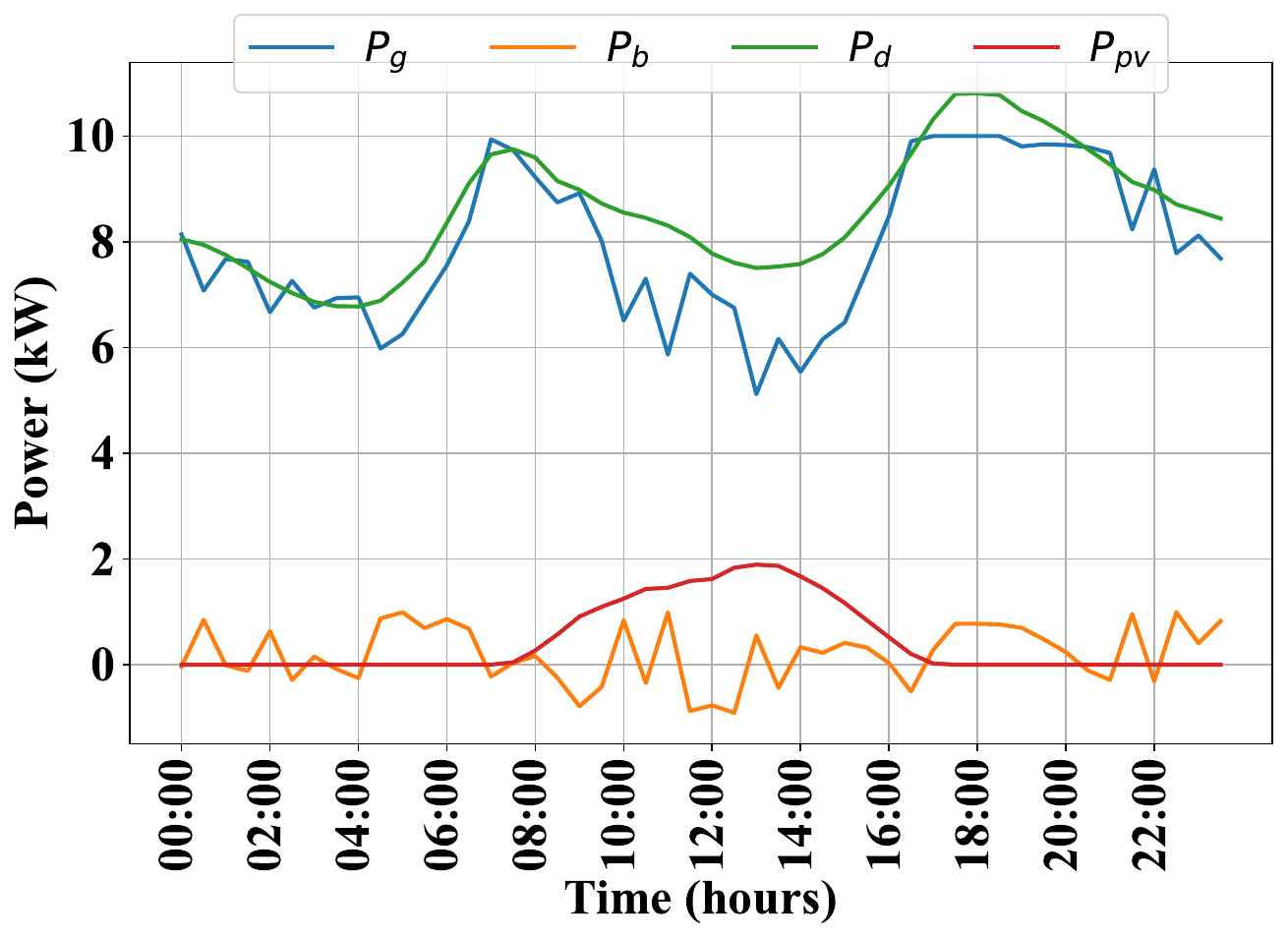}}
  \hfil
  \subfloat[]{\includegraphics[width=0.33\textwidth]{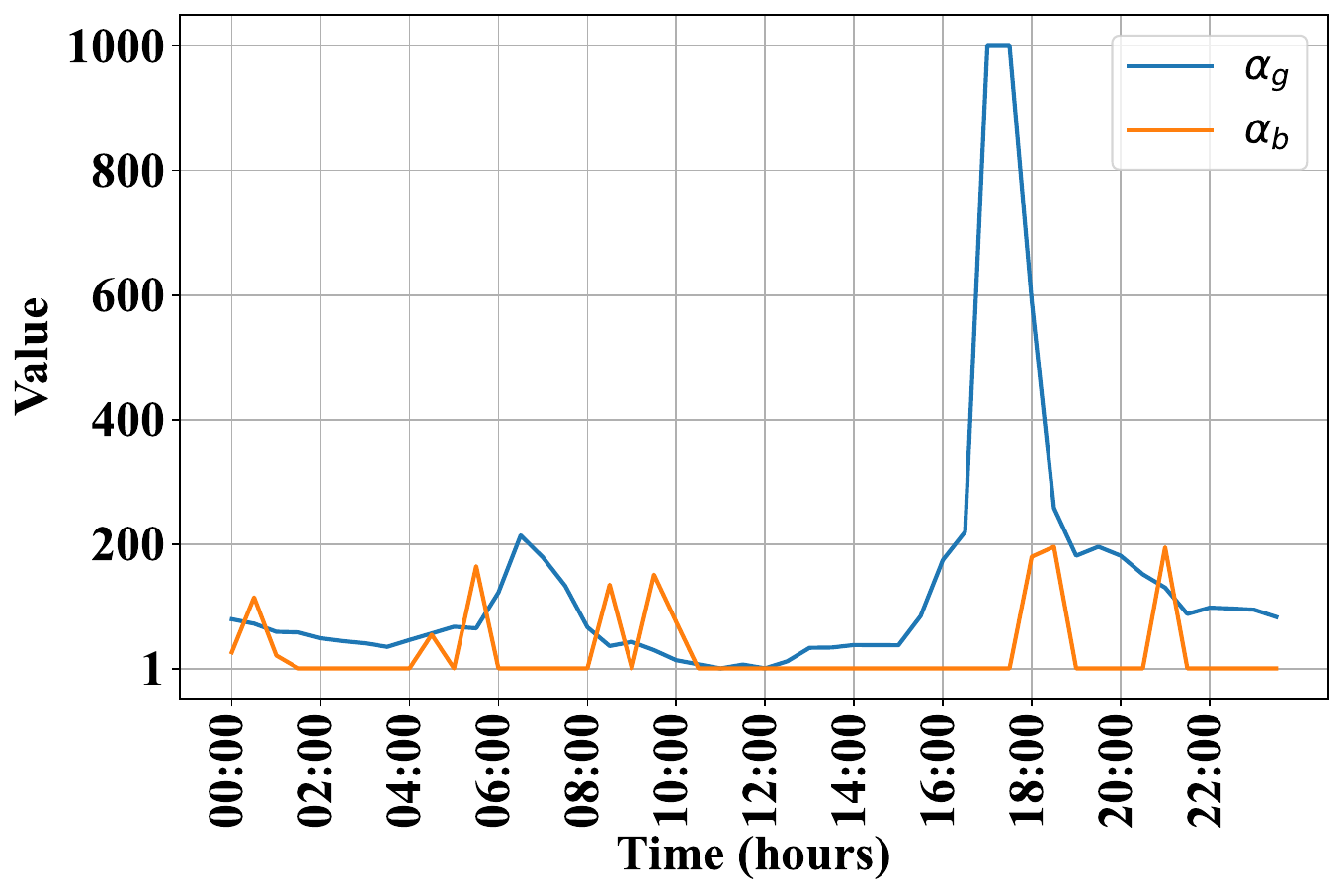}}
  \hfil
  \subfloat[]{\includegraphics[width=0.325\textwidth]{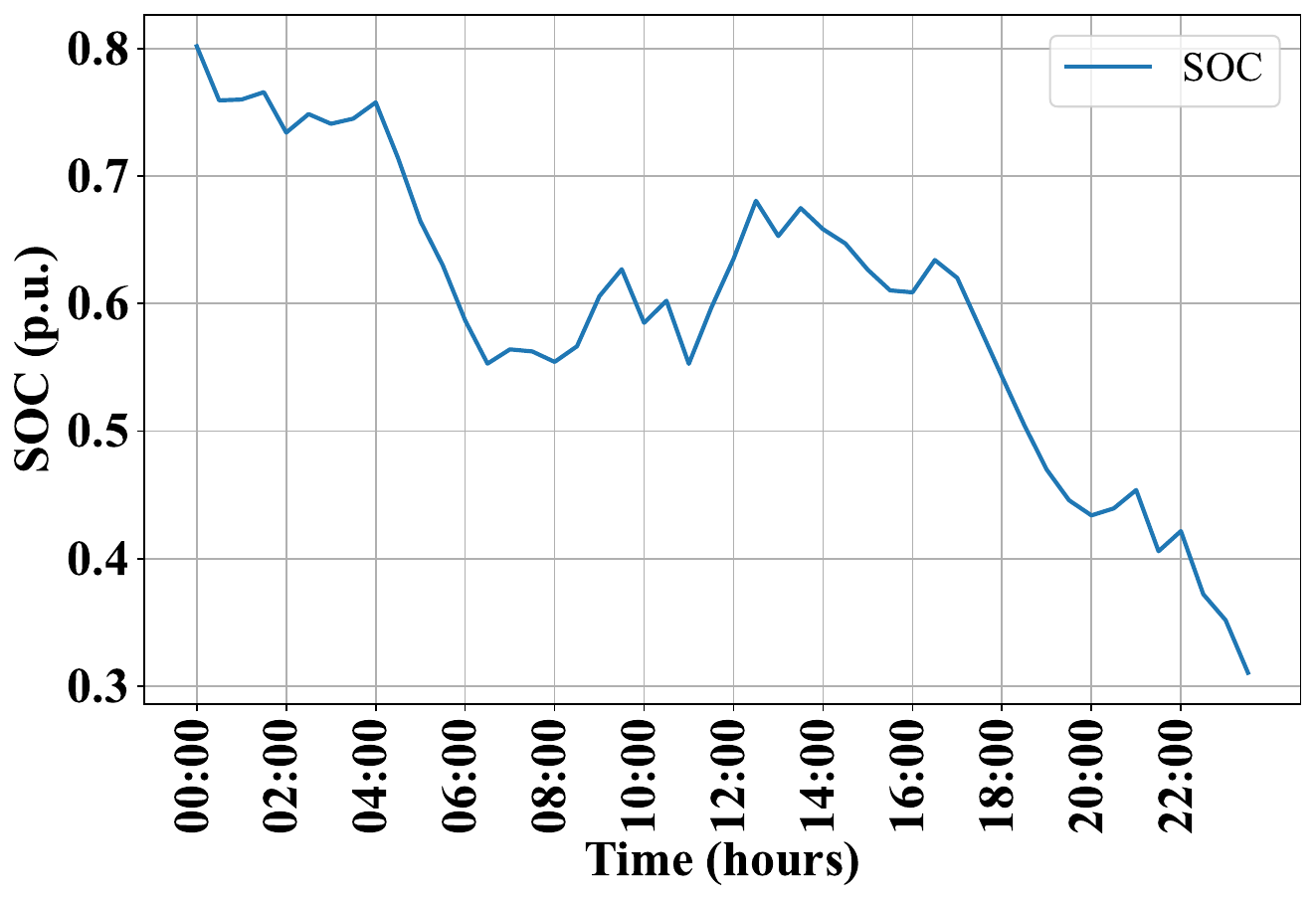}}

  \raisebox{-0.9\totalheight}[0pt][0pt]{\rotatebox{90}{QLD}}
  \subfloat[]{\includegraphics[width=0.32\textwidth]{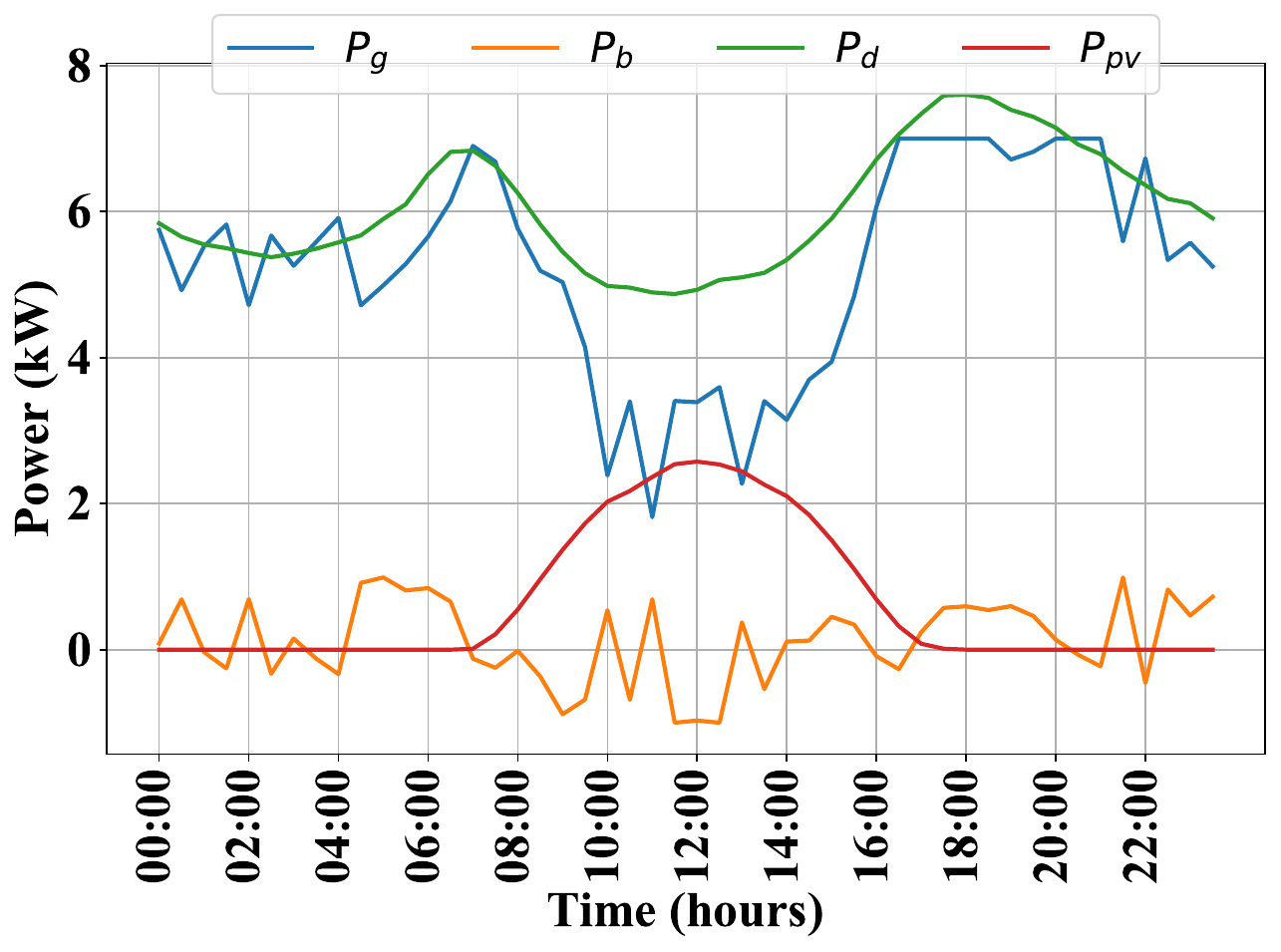}}
  \hfil
  \subfloat[]{\includegraphics[width=0.33\textwidth]{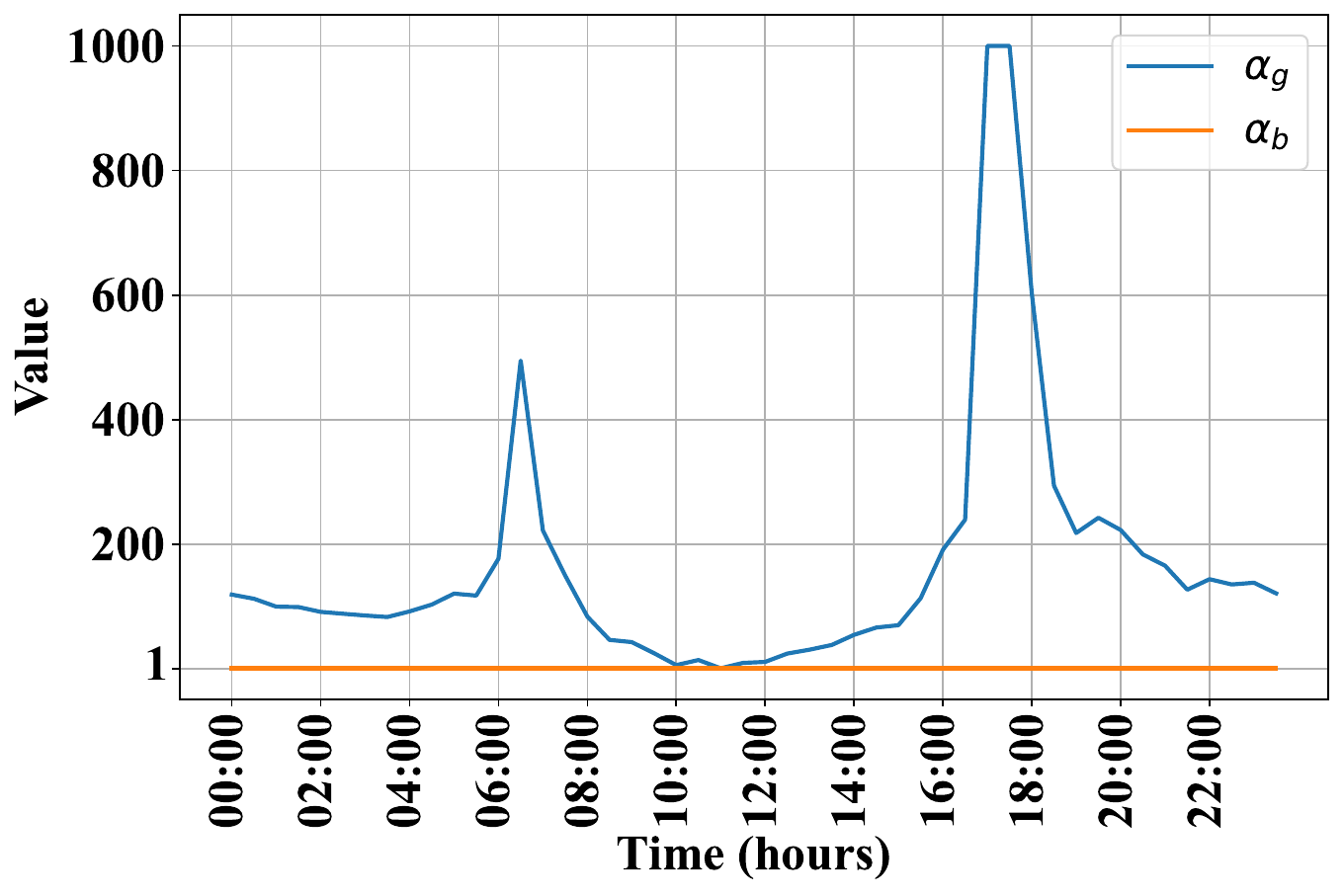}}
  \hfil
  \subfloat[]{\includegraphics[width=0.32\textwidth]{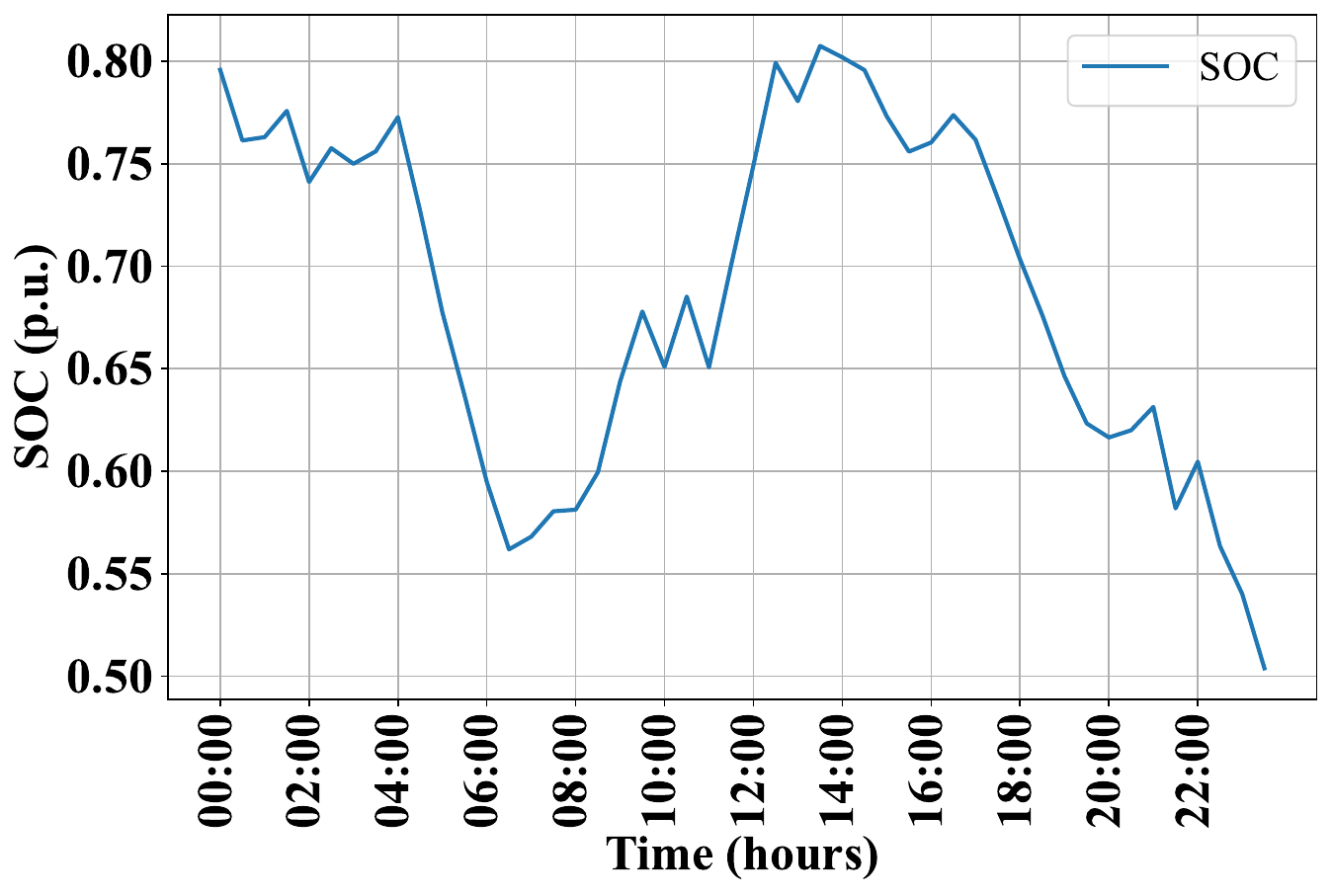}}

  \raisebox{-0.9\totalheight}[0pt][0pt]{\rotatebox{90}{SA}}
  \subfloat[]{\includegraphics[width=0.32\textwidth]{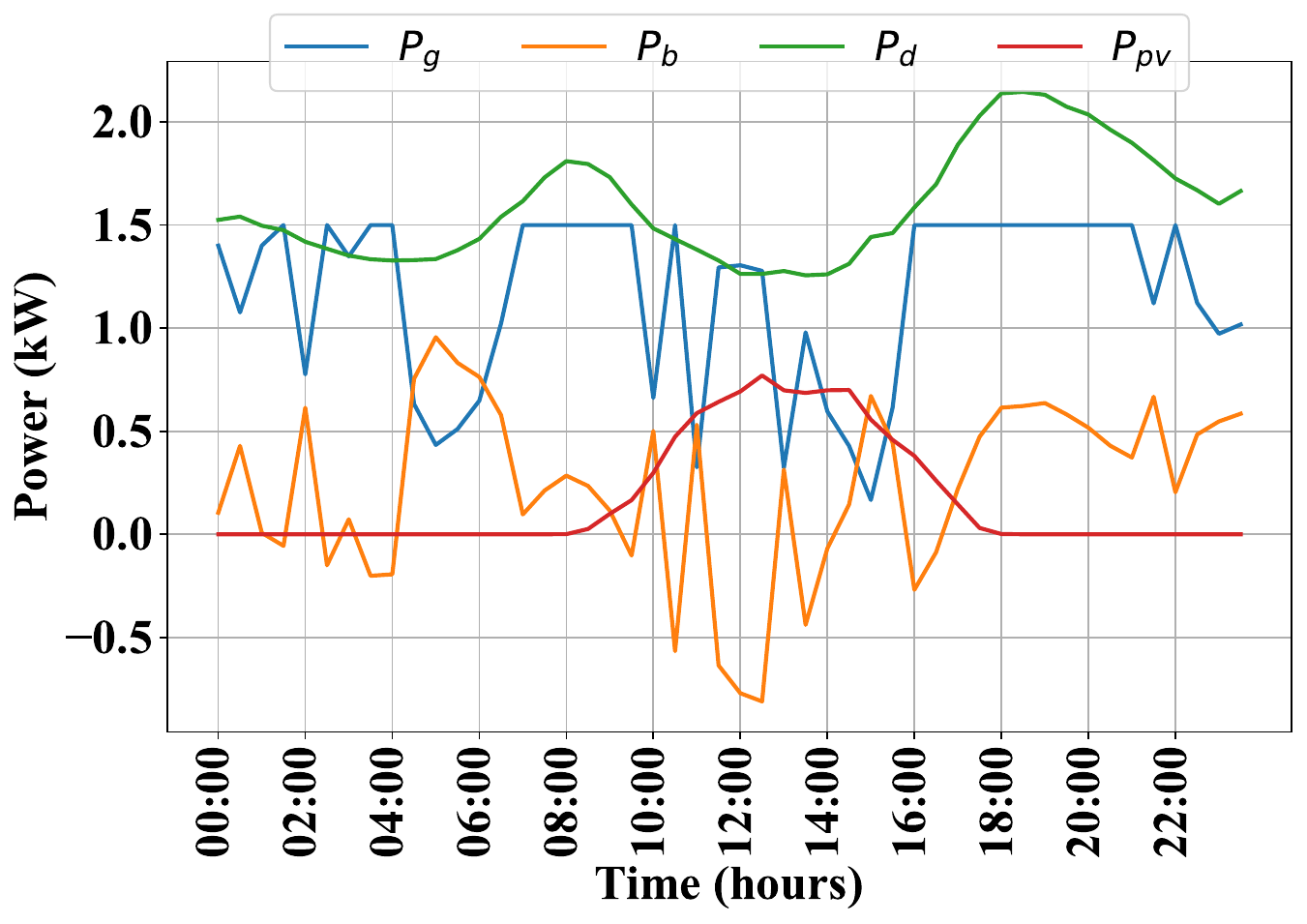}}
  \hfil
  \subfloat[]{\includegraphics[width=0.32\textwidth]{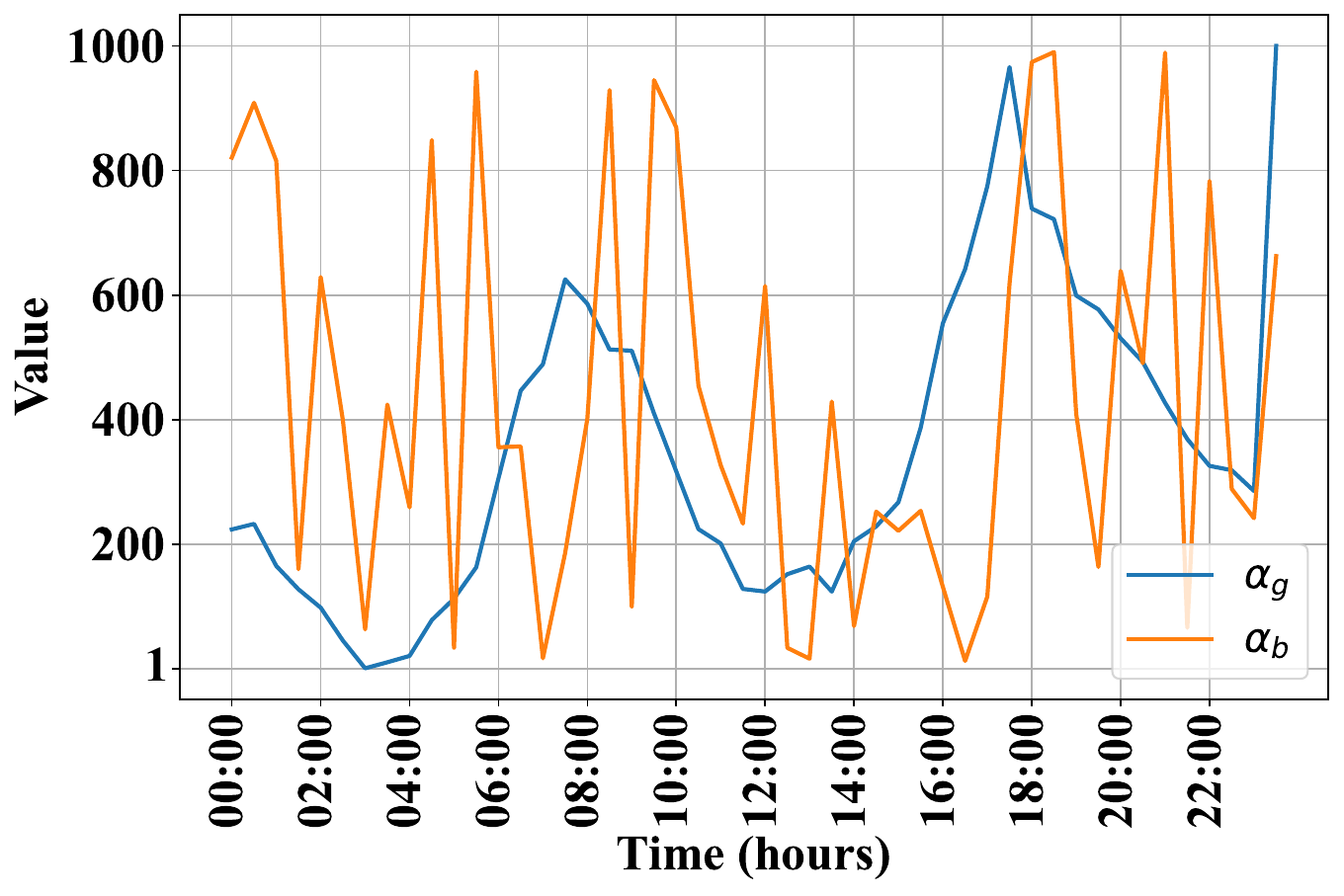}}
  \hfil
  \subfloat[]{\includegraphics[width=0.32\textwidth]{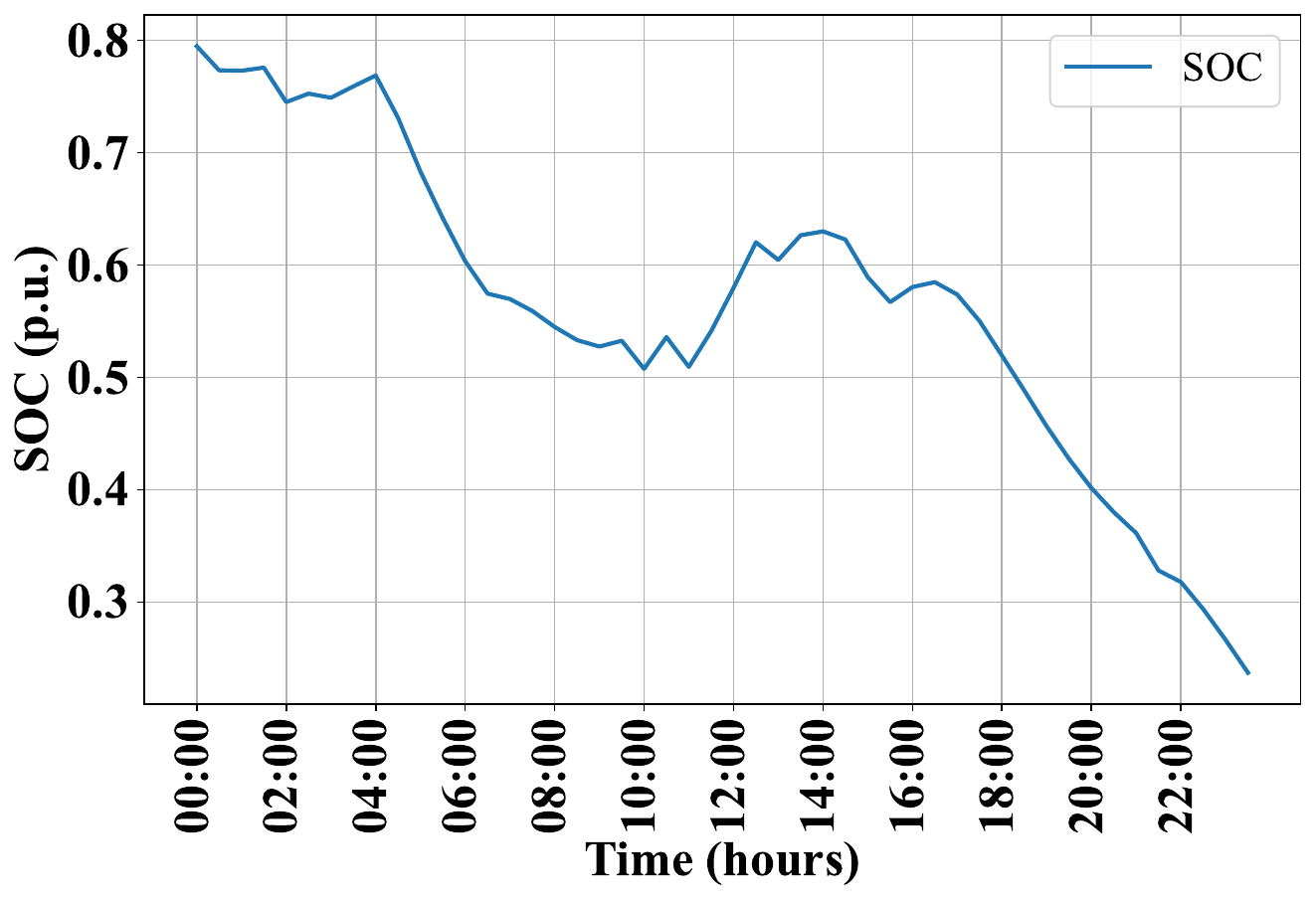}}

  \raisebox{-0.9\totalheight}[0pt][0pt]{\rotatebox{90}{TAS}}
  \subfloat[]{\includegraphics[width=0.32\textwidth]{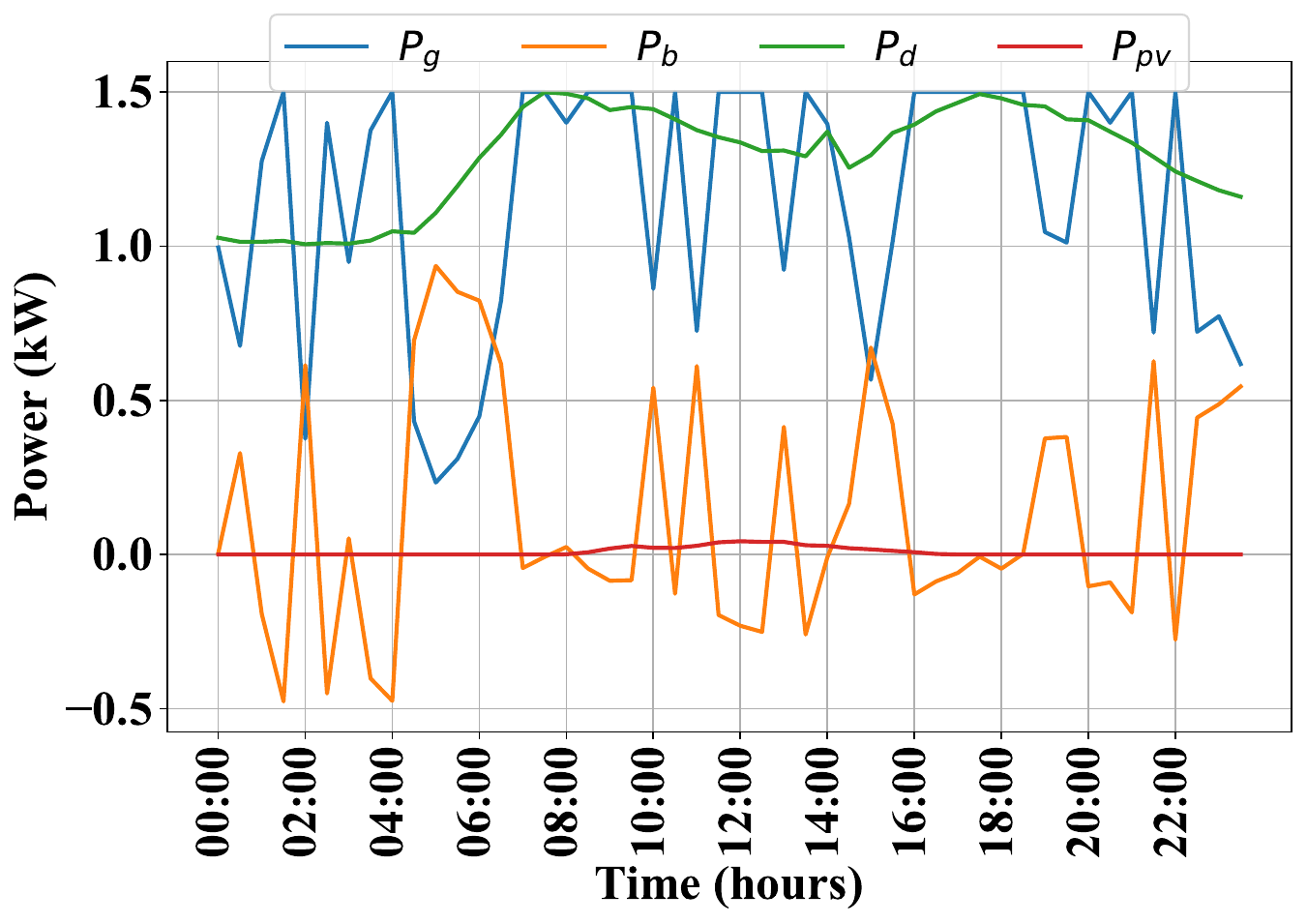}}
  \hfil
  \subfloat[]{\includegraphics[width=0.32\textwidth]{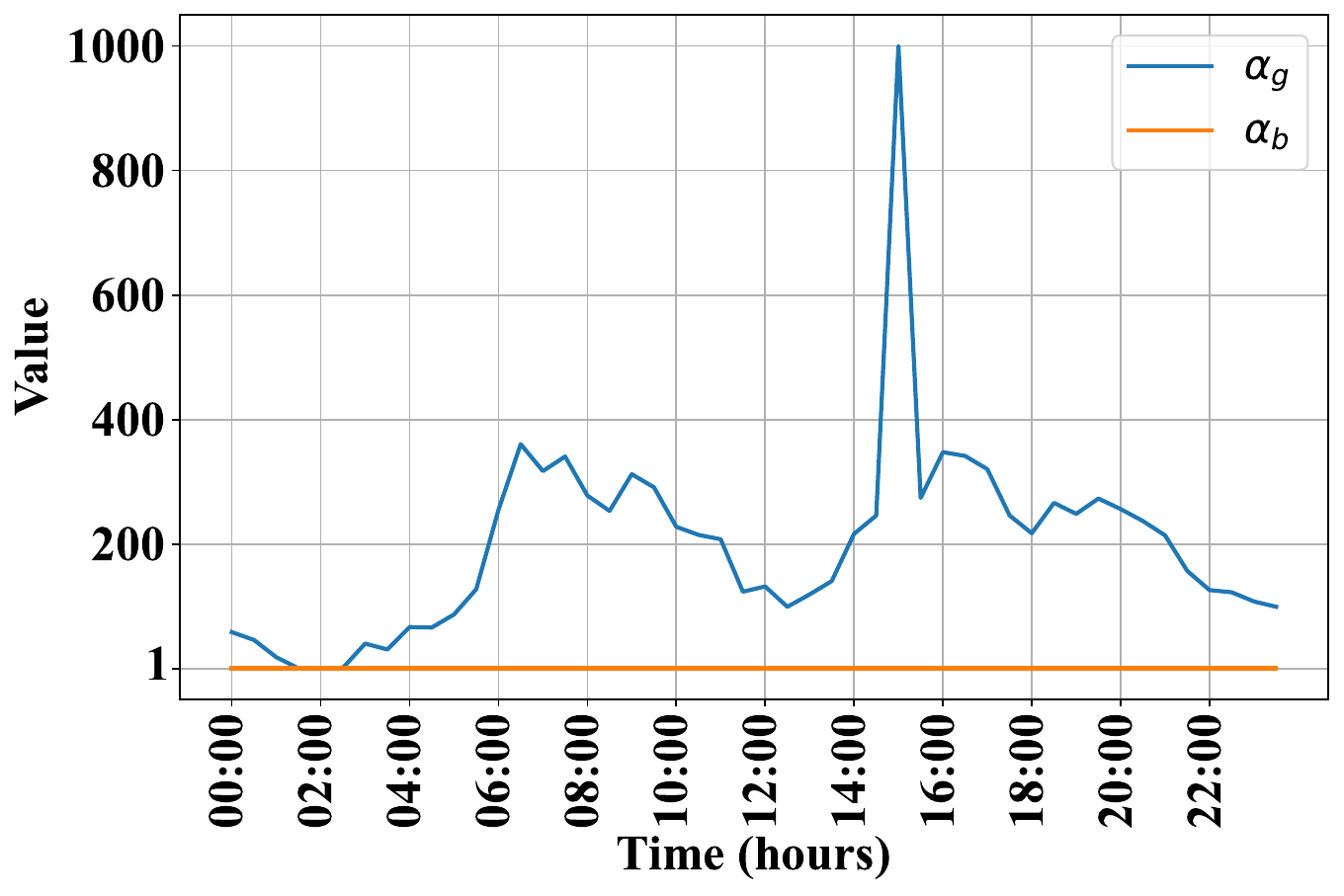}}
  \hfil
  \subfloat[]{\includegraphics[width=0.32\textwidth]{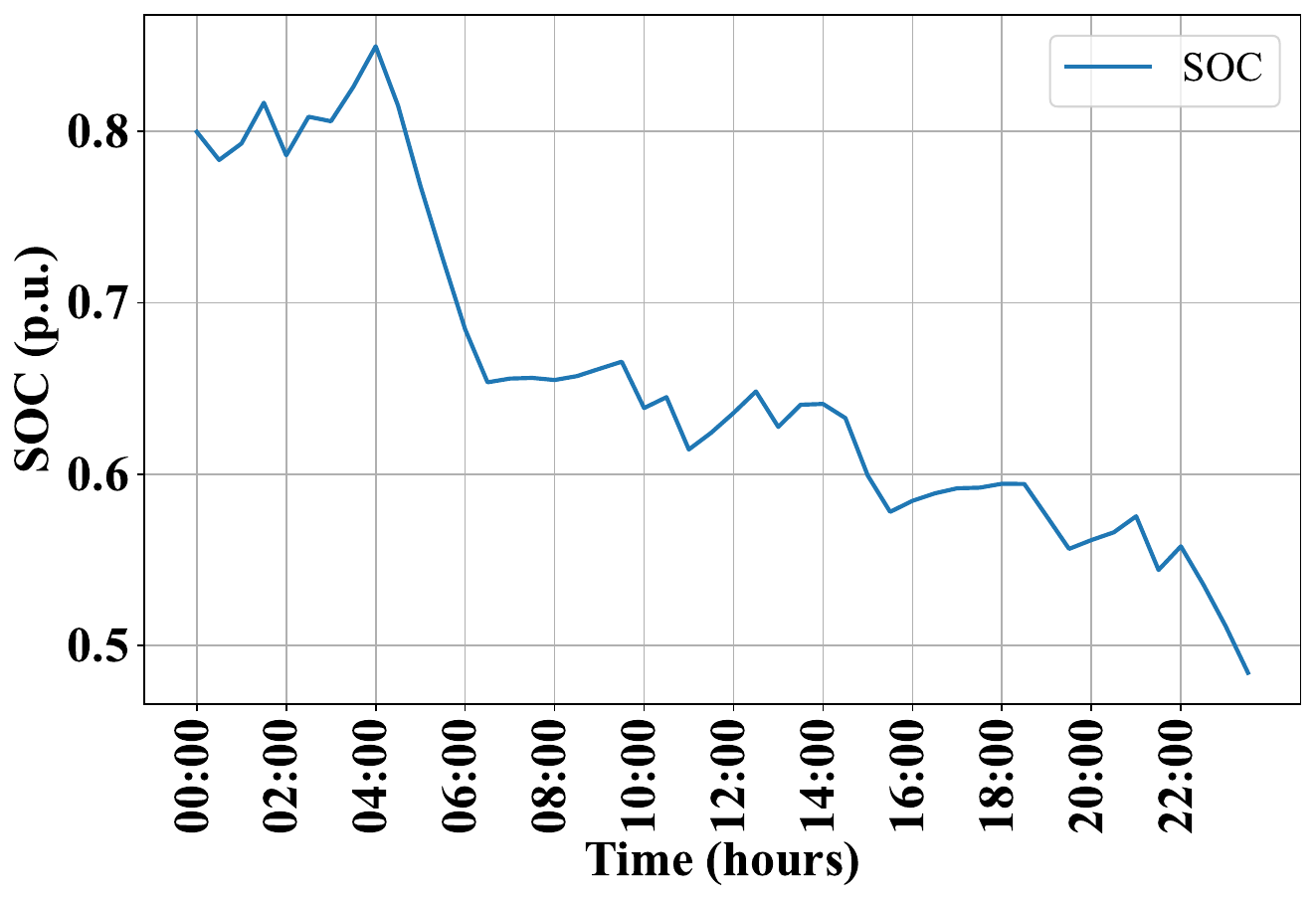}}

  \raisebox{-0.9\totalheight}[0pt][0pt]{\rotatebox{90}{VIC}}
  \subfloat[]{\includegraphics[width=0.305\textwidth]{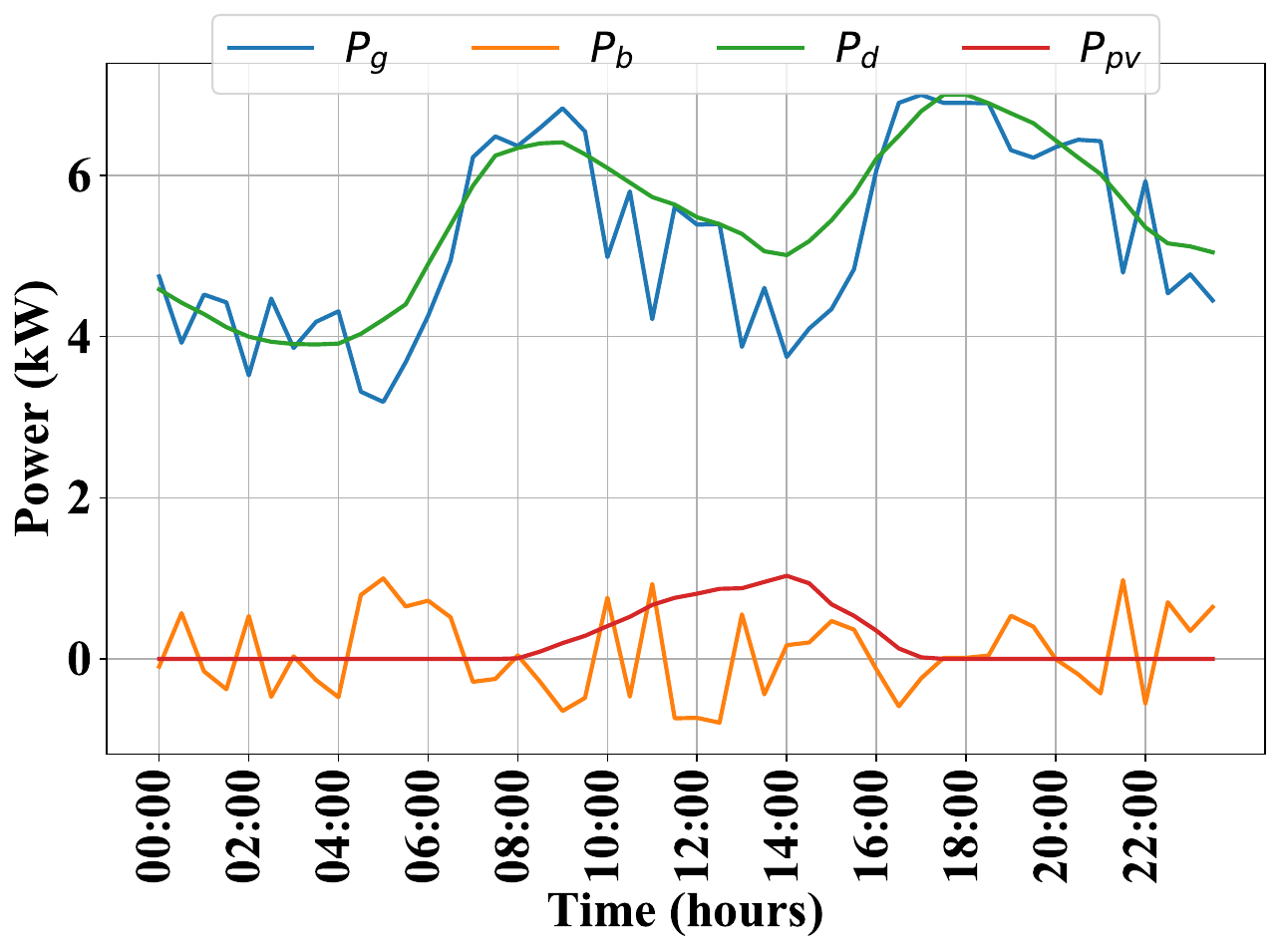}}
  \hfil
  \subfloat[]{\includegraphics[width=0.305\textwidth]{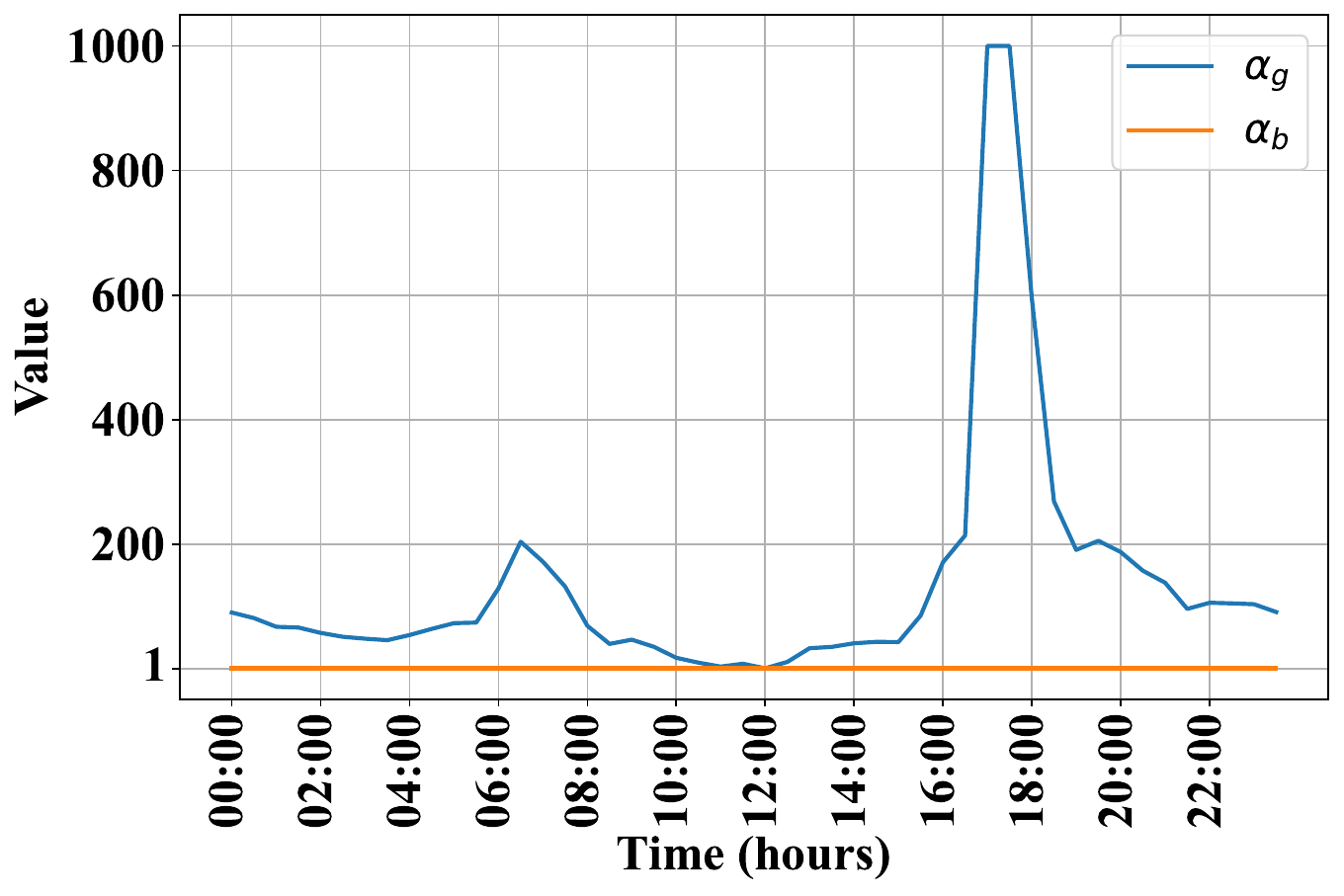}}
  \hfil
  \subfloat[]{\includegraphics[width=0.305\textwidth]{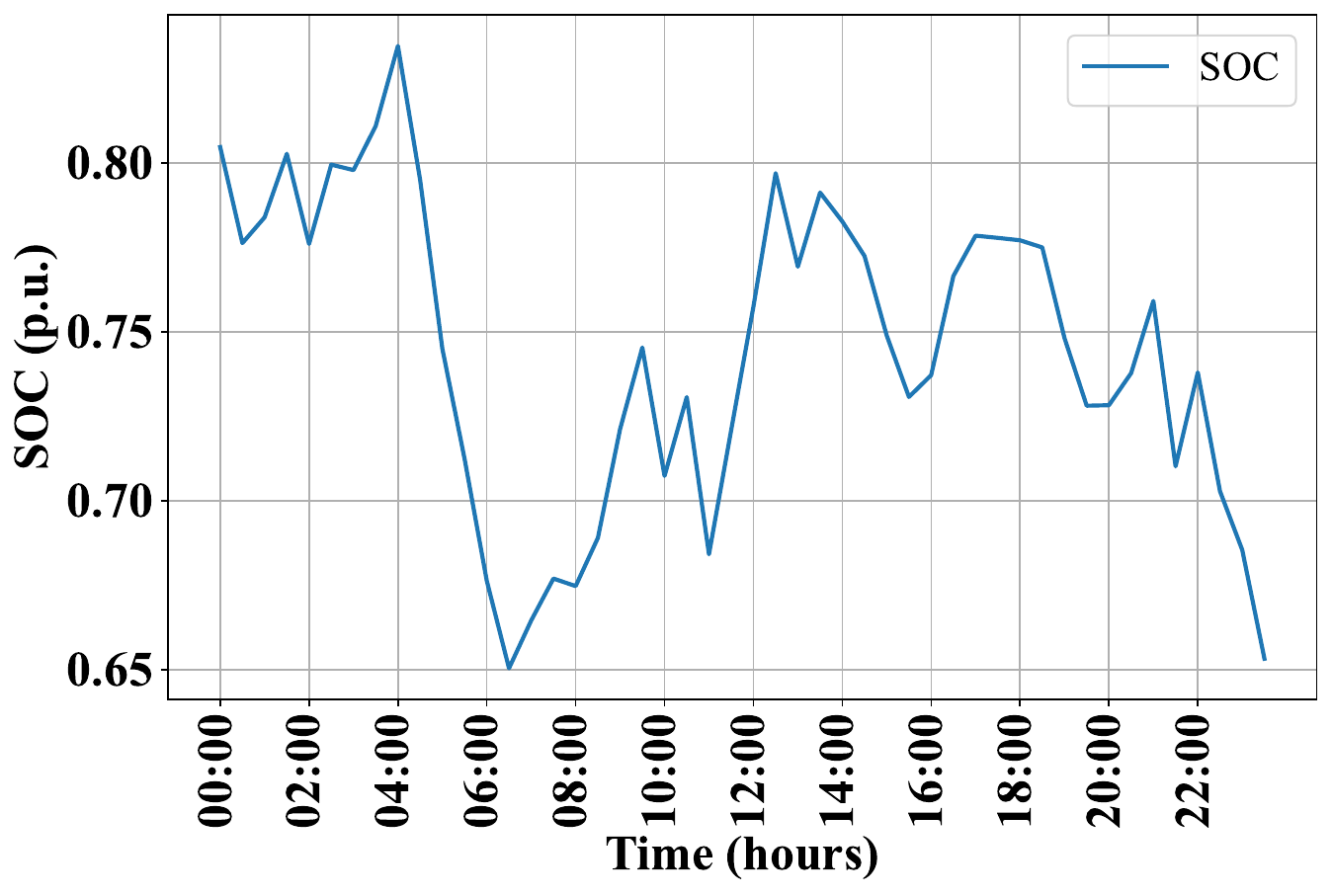}}

  \caption{Samples generated by the learning-based algorithm for the optimal BESS pattern for Australian states' districts.}
  \label{fig:subplots}
\end{figure*}

\begin{table*}[t]
\centering
\caption{Optimization results for gradient-based approach}
\label{tab:optimization_results}
\begin{tabular}{cccccc}
\hline
\textbf{Region} & \textbf{Converged after iterations} & \textbf{Total cost (\$)}  & \textbf{SOC at the end of the period (\%)} & \textbf{Total optimization time (s)} \\
\hline
NSW & 103100 & 495.49  & 0.25 & 2336.17 \\
QLD & 121300 & 498.27  & 0.40 & 3985.19 \\
SA & 90600 & 216.21  & 0.19 & 3045.87 \\
TAS & 104000 & 111.15  & 0.39 & 3731.02 \\
VIC & 106400 & 397.99  & 0.52 & 2620.54 \\
\hline
\end{tabular}
\end{table*}



Numerical experiments are conducted on a 64-bit machine equipped with an Intel Core i9-12900KF CPU running at 3.19GHz, 128GB of RAM. Through multiple iterations, the computational complexity and convergence behavior of the optimization approach are examined. 
It is found that setting the number of epochs at 5,000,000 is the most favorable for ensuring convergence. Additionally, a targeted hyper-parameter tuning experiment is implemented to ascertain the most effective learning rate and decay rate, leading to the adoption of a learning rate of 0.1 and a decay rate of 0.95. These parameters culminated in convergence after approximately 98,800 iterations in each tuning scenario. Also, we consider here the uniform distribution for the system uncertainties, and other distributions will be discussed later.

Table \ref{tab:optimization_results} and Fig.\ref{fig:subplots} present the optimization results for various regions using a fixed number of 5,000,000 epochs. It is noticeable that the number of iterations required for convergence varies between regions. For instance, the optimization for SA converges relatively faster, at 90,600 iterations, compared to QLD, which takes 121,300 iterations. This could be attributed to the characteristics and complexity of the load profiles in different regions. Regarding the total electricity cost for one day, TAS has the lowest at \$111.15, while QLD records the highest at \$498.27. These costs reflect the effectiveness of the optimization process in minimizing expenses associated with grid interactions and possibly the region-specific characteristics like electricity prices or load patterns. It is commendable that there are no SOC violations across all regions, indicating that the constraints on SOC are successfully adhered to throughout the optimization process. However, the SOC at the end of the period varies significantly among the regions. SA has the lowest final SOC percentage at 0.19, whereas TAS has the highest at 0.39. These SOC percentages could indicate the strategy used in energy storage and dispatch for each region and how the optimization algorithm has tailored itself to meet region-specific requirements. Remarkably, the SOC is discovered to fall below the preferred 50\% threshold at the end of the period, indicating unfavorable conditions for the battery's use the following day. This observation underscores the importance of considering the final SOC in the optimization process and motivates the next steps in our research framework. Having identified a set of feasible solutions, our subsequent aim is to refine these solutions and ascertain the most optimal among them, particularly aligning the final SOC with the set target.

Additionally, we control here the $\alpha$ parameters to optimize the performance of the BESS for the scheduling problem. Specifically, $\alpha_g$ is adjusted to become significantly high during periods of high tariff prices. This imposes a stronger penalty on grid dispatch, effectively discouraging the use of grid power when it is most expensive. Conversely, $\alpha_b$ is increased during specific times when there is a concern about excessive battery power dispatch. This prevents the batteries from depleting their SOC below 50\%, ensuring that they remain sufficiently charged for future use. This strategic adjustment of $\alpha$ parameters helps to optimize BESS performance and avoid costly dispatch scenarios, as evidenced in the results for the SA case. Furthermore, the total optimization time showcases the computational demand of this method. QLD consumes the most time at approximately 3,985 seconds, whereas NSW is the quickest at around 2,336 seconds. This suggests that, despite setting a uniform number of epochs, the computational complexity and convergence speed of the optimization process are influenced by the inherent characteristics of the regions' load profiles and possibly the initial conditions.

\subsection{Deep Reinforcement Learning Approach}

\begin{table*}[t]
  \centering
  \caption{Proposed DRL agent testing results for real load profiles}
  \label{tab:data}
  \begin{tabular}{cccccc}
    \toprule
    \textbf{State} & \textbf{Training Time (s)} & \textbf{Cumulative Grid Cost (\$)} & \textbf{Minimum SOC reached (\%)} & \textbf{SOC at the end of the period (\%)} \\
    \midrule
    NSW & 1653.96 & 367.93 & 0.64 & 0.74 \\
    QLD & 1602.69 & 320.25 & 0.59 & 0.67 \\
    SA & 1948.21 & 241.26  & 0.51 & 0.45 \\
    TAS & 1554.86  & 120.54 & 0.58 & 0.81  \\
    VIC & 2001.33 & 332.41 & 0.57 & 0.61  \\
    \bottomrule
  \end{tabular}
\end{table*}

\begin{figure*}[h]
\centering
\includegraphics[width=1.0\textwidth]{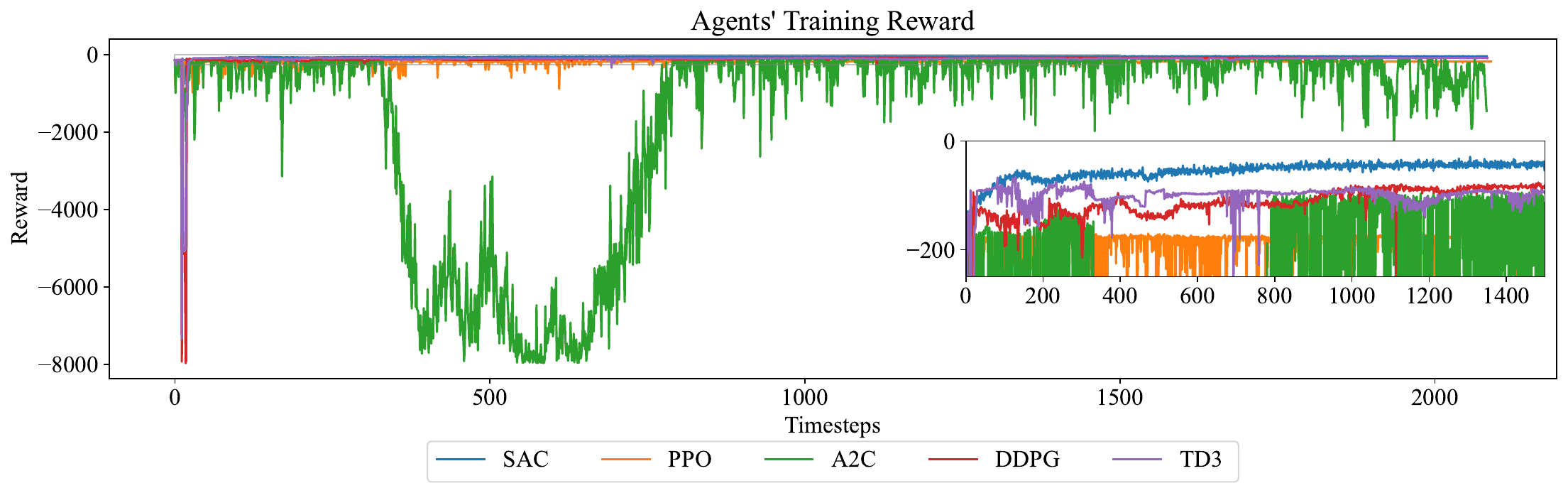}
\centering
\caption{DRL agents training results\label{DRL agents training results}}
\end{figure*}

\begin{figure}[!ht]
\centering
\includegraphics[width=1.0\columnwidth]{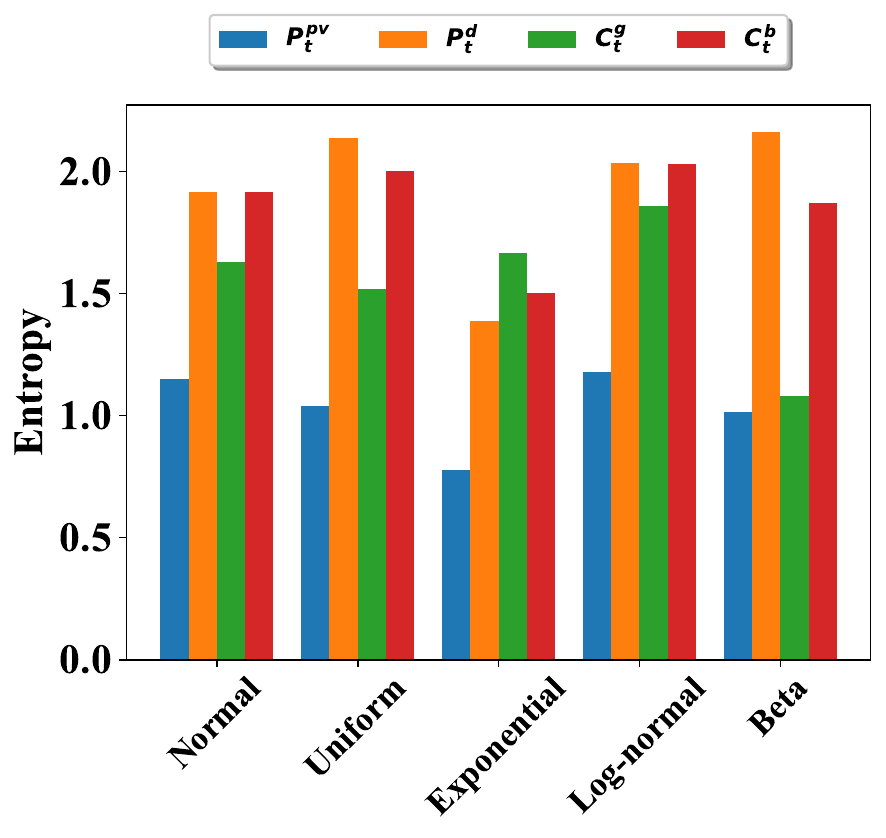}
\centering
\caption{Entropy analysis for uncertainty distributions\label{DRL agents training results entropy}}
\end{figure}

\begin{figure}[!ht]
\centering
\includegraphics[width=1.0\columnwidth]{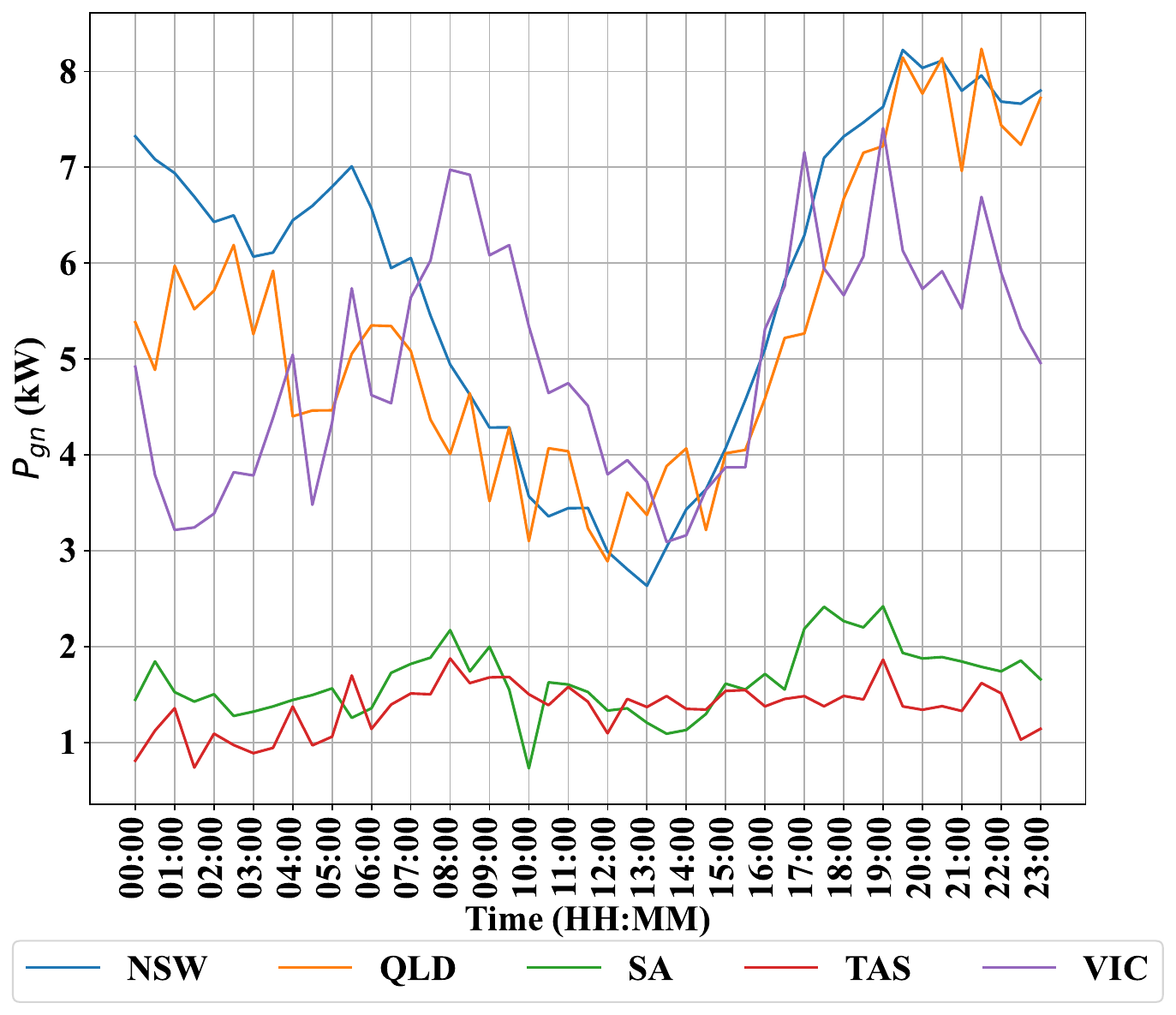}
\centering
\caption{SAC testing results for the grid power set points \label{Pg_plot}}
\end{figure}

\begin{figure}[!ht]
\centering
\includegraphics[width=1.0\columnwidth]{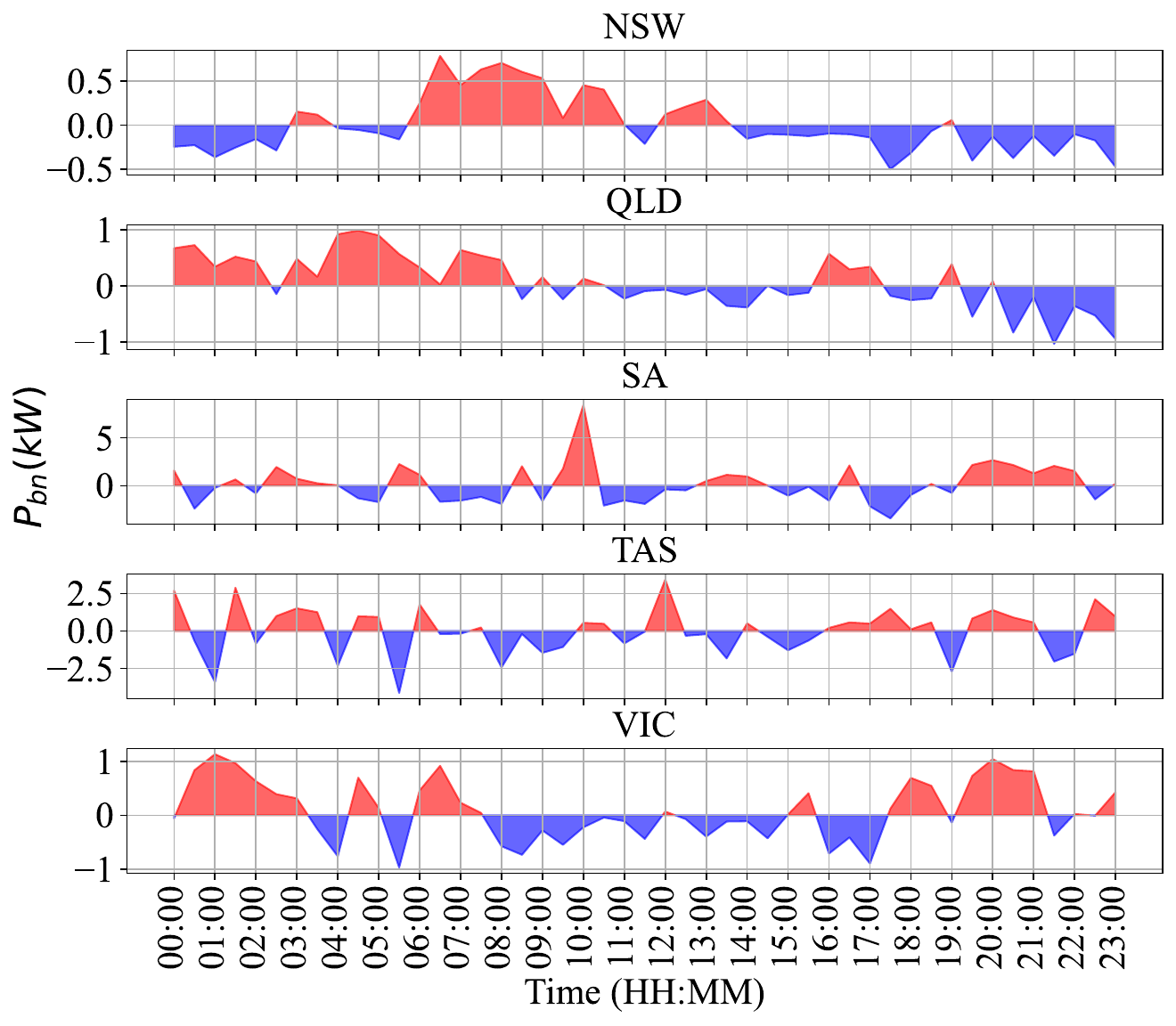}
\centering
\caption{SAC testing results for the battery power set points\label{Pb_subplots}}
\end{figure}

\begin{figure}[!ht]
\centering
\includegraphics[width=1.0\columnwidth]{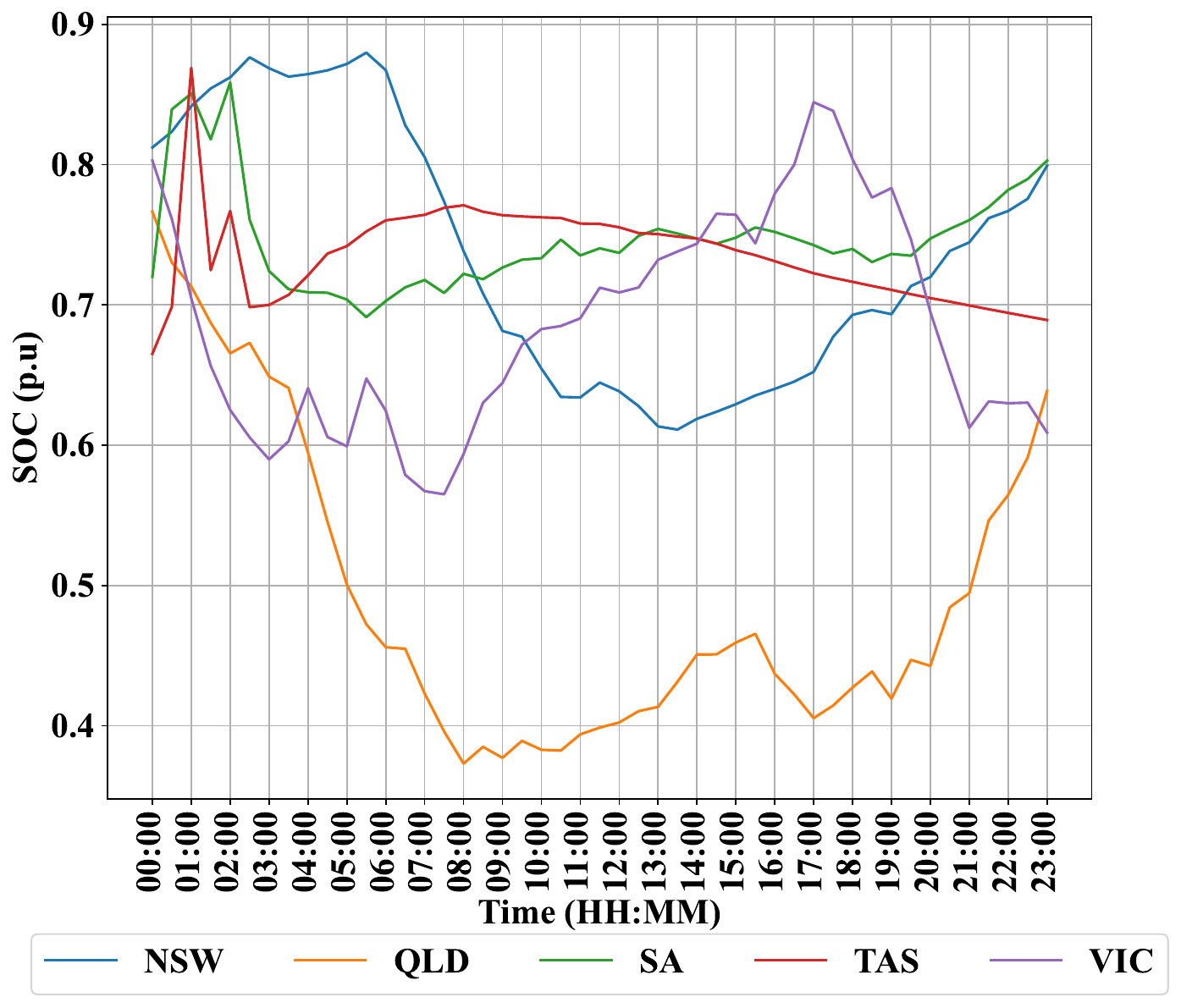}
\centering
\caption{SAC testing results for SOC profiles \label{SOC_plot}}
\end{figure}


The same PC setup used in a gradient-based method for testing is employed in this study. We conduct a comparative evaluation of five DRL algorithms: SAC, PPO, A2C, DDPG, and TD3. SAC exhibits the most favorable performance based on the learning curve, as depicted in Fig.\ref{DRL agents training results}. It achieves the fastest convergence toward the global optimal reward value, effectively balancing the power supply and demand. Furthermore, the policy derived from SAC does not violate any system constraints, which demonstrates the robustness and reliability of this approach in maintaining the stability of the BESS. The ability of SAC to manage uncertainties in the power generated by photovoltaic cells and the power demand further attests to its robustness. The DDPG and TD3 algorithms also demonstrate rapid convergence toward the optimal solution. However, in contrast to SAC, these methods exhibit some instability during the training process. This results in occasional violations of the system constraints during testing and a failure to consistently achieve the optimal set point for controlling the decision variables. Therefore, while DDPG and TD3 offer advantages in terms of convergence speed, they may be less reliable for managing a BESS in real-world conditions, where adherence to system constraints is of paramount importance. Finally, the A2C and PPO algorithms exhibit poorer performance. These methods are slower to converge toward the optimal solution, and frequently violate the system constraints. Moreover, they generally fail to achieve the optimal set point for controlling the decision variables. These findings suggest that A2C and PPO may be ill-suited for the task at hand. In our comprehensive exploration of uncertainty across different probability distributions, we conduct detailed calculations of entropy for each distribution as shown in Fig.\ref{DRL agents training results entropy}. This process allows us to measure the inherent randomness embedded in each scenario, thereby assessing our models' ability to manage various levels of uncertainty. We examine five types of distributions: Normal, Uniform, Exponential, Log-normal, and Beta, and each is applied to four unique variables: $P_t^{pv}$, $P_t^d$, $C_t^g$, and $C_t^b$. In our model, the values assigned to the distributions are specifically selected to reflect real-time operation behavior. The normal distribution \( \mathcal{N}(\mu, \sigma) \) uses a mean \( \mu \) of 0 and a standard deviation \( \sigma \) of 0.10. The uniform distribution \( \mathcal{U}(a, b) \) ranges from \(-0.10\) to \(0.10\). The exponential distribution \( \mathcal{E}(\lambda) \) has a rate \( \lambda \) of 0.10, with an additional shift of \(-0.10\) to accommodate operational adjustments. The log-normal distribution \( \mathcal{LN}(\mu, \sigma) \) features a location \( \mu \) of 0 and a scale \( \sigma \) of 0.10, adjusted by \(-1.10\). Finally, the beta distribution \( \mathcal{B}(\alpha, \beta) \) employs shape parameters \( \alpha \) and \( \beta \) both set at 2, scaled by \(0.10 \cdot x_i\) and shifted by \(-0.5\), fine-tuned for real-time dynamics. Consequently, the entropy values associated with the $P_t^{pv}$ variable showed a considerable range of variation, swinging from a low of 0.78 in the Exponential distribution to a high of 1.18 in the Log-normal distribution. These figures underscore the substantial degree to which the randomness linked with $P_t^{pv}$ could fluctuate, depending on the selected distribution. A similar pattern is also evident in the entropy values for the other three variables, $P_t^d$, $C_t^g$, and $C_t^b$. Each exhibited significant variance depending on the distribution: $P_t^d$ entropy ranged from 1.38 (Exponential) to 2.16 (Beta), $C_t^g$ from 1.08 (Beta) to 1.86 (Log-normal), and $C_t^b$ from 1.50 (Exponential) to 2.03 (Log-normal). 
\begin{table*}[h!]
\centering
\small
\caption{For each distribution (Uniform, Beta, Gauss, Exp., Log) and each state solved by using two different approaches: Gradient Based and DRL. \textquotesingle 1\textquotesingle\  refers to end state of charge (\%) and \textquotesingle 2\textquotesingle\ refers to optimization time (seconds).}
\resizebox{\textwidth}{!}{%
\begin{tabular}{|l|c|c|c|c|c|c|c|c|c|c|c|c|c|c|c|c|c|c|c|c|}
\hline
& \multicolumn{10}{c|}{\textbf{Gradient-based}} & \multicolumn{10}{c|}{\textbf{DRL}} \\
\cline{2-21}
\textbf{State} & \multicolumn{2}{c|}{\textbf{Uniform}} & \multicolumn{2}{c|}{\textbf{Beta}} & \multicolumn{2}{c|}{\textbf{Gauss}} & \multicolumn{2}{c|}{\textbf{Exp.}} & \multicolumn{2}{c|}{\textbf{Log}} & \multicolumn{2}{c|}{\textbf{Uniform}} & \multicolumn{2}{c|}{\textbf{Beta}} & \multicolumn{2}{c|}{\textbf{Gauss}} & \multicolumn{2}{c|}{\textbf{Exp.}} & \multicolumn{2}{c|}{\textbf{Log}} \\
\cline{2-21}
& 1 & 2 & 1 & 2 & 1 & 2 & 1 & 2 & 1 & 2 & 1 & 2 & 1 & 2 & 1 & 2 & 1 & 2 & 1 & 2 \\
\hline
NSW & 0.25 & 4563 & 0.11 & 4351 & 0 & 8927 & 0.25 & 4382 & 0 & 4578 & 0.74 & 1525 & 0.53 & 2866 & 0.99 & 6207 & 0.75 & 2883 & 1.0 & 2884 \\
QLD & 0.40 & 5415 & 0.28 & 4511 & 0.04 & 7851 & 0.4 & 5310 & 0.04 & 4675 & 0.67 & 1603 & 0.76 & 2901 & 0.09 & 2731 & 0.60 & 2875 & 0.40 & 2896 \\
SA & 0.19 & 4062 & 0.17 & 4335 & 0.02 & 4707 & 0.19 & 3866 & 0.22 & 4313 & 0.51 & 1948 & 1.0 & 2048 & 1.0 & 2670 & 1.0 & 2948 & 0.05 & 3018 \\
TAS & 0.39 & 4648 & 0.34 & 4481 & 0.22 & 5085 & 0.02 & 4275 & 0.22 & 4313 & 0.81 & 1555 & 1.0 & 2400 & 0.73 & 3353 & 1.0 & 2916 & 0.96 & 2934 \\
VIC & 0.52 & 4757 & 0.37 & 4330 & 0.52 & 5204 & 0.52 & 4583 & 0.11 & 4588 & 0.61 & 2001 & 0.27 & 2949 & 0.66 & 2492 & 0.72 & 3058 & 0.63 & 3035 \\
\hline
\end{tabular}
}
\label{table:1}
\end{table*}

The testing results of the proposed SAC agent shown in Fig. \ref{Pg_plot} to Fig.\ref{SOC_plot} and Table\ref{tab:data} across including uniform distribution of uncertainties reveal that the DRL agent has been highly effective in reducing the cumulative grid costs in energy management. Note that this analysis focuses on the relative reduction in grid costs in comparison to gradient-based solutions and does not compare costs between states, as different regions might have varying energy tariffs and grid conditions. In the context of relative performance against gradient-based solutions, the significant reductions in cumulative grid costs as observed in the results for all states are indicative of the DRL agent's ability to efficiently manage energy storage systems. The agent appears to have learned to make more informed and optimized decisions in real-time, which, in contrast to gradient-based solutions, allows it to adapt better to the variability and dynamics of the load profiles. Additionally, the minimum SOC reached by the storage systems during the testing period is relatively low across all states. This suggests that the DRL agent effectively utilizes the storage capacity to minimize reliance on the grid, possibly more effectively than gradient-based methods, which might not fully exploit the available storage capacity. Furthermore, the SOC at the end of the testing period also varies, with values suggesting that the DRL agent's policies effectively manage the energy storage, ensuring that there is enough charge retained or consumed based on the requirement. Moreover, the minimum SOC reached by the battery storage systems during the testing period is relatively low across all states, with the lowest being 0.51\% in SA. This indicates that the agent effectively utilizing the storage capacity to minimize grid reliance, though care must be taken to ensure that such low SOC levels do not compromise the lifespan of the storage systems. Finally, the SOC at the end of the testing period is between 0.45\% in SA and 0.81\% in TAS. This indicates that the DRL agent's policies tend to deplete the storage to some extent by the end of the period. However, the significantly higher SOC in TAS suggests a more conservative approach or an abundance in energy availability during that period.

Table \ref{table:1} provides a comprehensive comparison between the gradient-based and DRL optimization approaches across different Australian states. This comparison is conducted based on two performance metrics - the end state of charge in percentage \textquotesingle 1\textquotesingle\ and optimization time in seconds \textquotesingle 2\textquotesingle\, for five different statistical distributions: Uniform, Beta, Gauss, Exponential, and Lognormal. Considering the gradient-based approach first, we observe that the results for the end state of charge and optimization time vary across states and distributions. For instance, NSW exhibits the lowest end SOC under the Gaussian distribution (0), and the lowest optimization time with the Beta distribution (4,351). On the contrary, QLD records the highest end SOC (0.4) under Exponential distribution, and the highest optimization time (7,851) in the Gauss distribution. TAS demonstrates a unique trend where the lowest end SOC (0.02) and the second lowest optimization time (4,275) are recorded under the Exponential distribution. Transitioning to the DRL approach, there are different patterns. TAS consistently achieves the highest end SOC across three distributions: Beta (1.0), Gauss (0.73), and Exponential (1.0). Conversely, NSW exhibits the most efficient optimization time under the Gauss distribution (6,207). The DRL approach achieves a perfect end SOC of 1.0 across multiple states and distributions - an achievement not seen with the gradient-based approach.
Comparing both methods, we see that the DRL approach tends to lead to a higher end state of charge across all states and distributions. For instance, it achieves an end SOC of 1.0 in SA under Beta distribution and in TAS under Beta and Exponential distributions, as opposed to the highest end SOC of 0.52 achieved by the gradient-based approach in VIC under the Gauss distribution. In terms of optimization time, the DRL approach also generally records lower times, with its highest being 6,207 in NSW under Gauss distribution, compared to the highest of 8,927 in NSW under Gauss distribution for the gradient-based approach. 

Examining the results in detail, it becomes evident that the DRL approach exhibits a stronger propensity for transferability, both spatially across states and temporally across distributions. This is particularly noticeable when observing the higher degree of consistency in achieving optimal or near-optimal end state of charge across multiple states and distributions. This implies that the DRL-based approach demonstrates a robust adaptability to changes in environment and conditions, thus indicating its enhanced transferability. For instance, the DRL approach reaches an end SOC of 1.0 across three distinct states - SA, TAS, and VIC, under various distributions - Beta, Gauss, and Exponential. This indicates that the DRL-based method can effectively adapt to different spatial regions with disparate conditions. Temporally, the DRL-based approach maintains relatively consistent optimization times across different distributions, suggesting that it could also handle temporal variations well. This robustness to spatial and temporal changes underscores the advantage of DRL, especially in real-world applications where conditions can be dynamic and unpredictable. It suggests that a DRL-based approach can be more readily transferred and scaled to other geographical regions and time-sensitive operations, providing a more flexible and reliable solution for optimizing battery storage systems under various conditions. This aspect of transferability is crucial for practical implementations, particularly in the context of future energy systems where adaptability to evolving circumstances will be paramount. The gradient-based approach, while generally performing well across the board, exhibits some inconsistencies in achieving optimal or near-optimal end states of charge. This suggests that while it can adapt to different distributions and states, its performance might not be as consistent or reliable as that of the DRL approach. The optimization time of the gradient-based approach also shows more variation, suggesting that it may be more sensitive to changes in the problem context. On the other hand, the DRL approach shows a greater degree of versatility. It not only reaches optimal or near-optimal end states of charge across multiple states and distributions but also maintains relatively consistent optimization times. This suggests that the DRL approach is capable of generalizing across different conditions more effectively and providing more consistent performance.

\subsection{Comparison of DRL and Gradient-based Methods}

\begin{figure}[!ht]
\centering
\includegraphics[width=1.0\columnwidth]{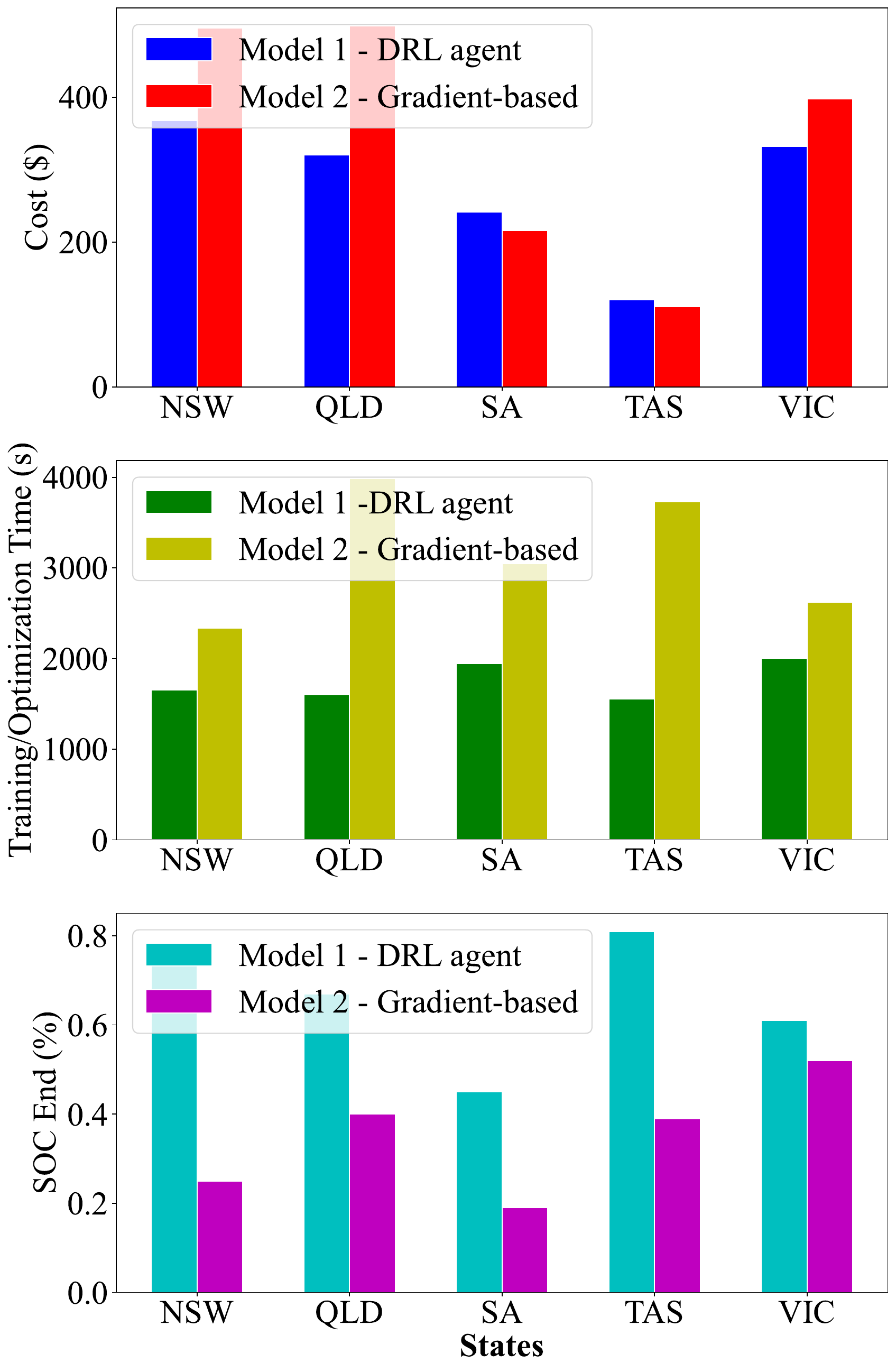}
\centering
\caption{Model comparisons\label{comparison_plot}}
\end{figure}

A comparative analysis of the results obtained from the DRL agent and the gradient-based optimization method reveals several insightful conclusions as shown in Fig.\ref{comparison_plot} regarding their performance in managing energy storage systems across different regions, as summarized in Table~\ref{tab:optimization_results} and Table~\ref{tab:data}.

\begin{itemize}
    \item \textbf{Cumulative Grid Costs:} One of the most striking observations is the substantial reduction in cumulative grid costs achieved by the DRL agent. For instance, in NSW, the DRL agent manages to reduce the costs to \$367.93, whereas the gradient-based method results in costs of \$495.49. This trend is consistent across other regions including QLD, and VIC, except in TAS and SA, where the cost evaluated by both methods are almost the same and indicating that the DRL agent is influenced by the spatial domain of data, presenting opportunities for future research aimed at improving its performance in spatially varying environments. 
    
    \item \textbf{State of Charge:} Another critical performance metric is the SOC at the end of the period. The DRL agent not only minimizes costs but also ensures a higher SOC in the energy storage systems. This is evident from the data; in NSW, for example, the DRL agent attains a SOC of 0.74 compared to a significantly lower 0.25 by the gradient-based method. 
    
    \item \textbf{Computational Efficiency:} The DRL agent’s training times are notably shorter than the total optimization times required by the gradient-based method. This aspect is vital for real-world implementations where timely decision-making is imperative. For instance, in NSW, the training time of the DRL agent is 1653.96 seconds, compared to 2336.17 seconds taken by the gradient-based method.
\end{itemize}

\section{Conclusion}

In this study, we addressed the imperative challenge of determining optimal power set points using a novel framework of RMFEMF, in the context of energy management. The main accomplishments and implications of our research are elaborated upon below:

\begin{enumerate}[label=(\roman*)]
    \item \textbf{Comparison between DRL Agents and Gradient-Based Methods for Set Point Determination:} We developed an innovative framework that introduced gradient-based optimization technique using Adam as a benchmark for intelligent and dynamic determination of $P_{g}$ and $P_{b}$ set points. By synergizing DRL’s learning capabilities with the fine-tuned optimization offered by hyper-parameters and reward design, the proposed framework exhibits enhanced performance and adaptability compared to traditional microgrid control strategies.
    
    \item \textbf{Strategic Identification and Utilization of Penalty Terms for Global Optimality:} A distinct element of our proposed framework is the systematic identification and integration of penalty terms within the optimization process. Through careful analysis, we discerned an optimal set of penalty terms essential for achieving globally optimal solutions for $P_{g}$ and $P_{b}$ set points.
    
    \item \textbf{Development and Employment of Comprehensive Evaluation Metrics:} To establish a robust evaluation of the proposed framework's performance, we devised a comprehensive set of evaluation criteria. These criteria include the SOC at the end of the period, SOC constraint violations, computational time required for training and optimization, and cumulative grid energy cost. These metrics collectively provided a holistic assessment, capturing both operational and economic facets of BESS management.
    
    \item \textbf{Robustness Validation through Comprehensive Testing:} To affirm the resilience and effectiveness of our proposed framework, we tested it across a multitude of uncertainty scenarios, adopting diverse distribution models for all input parameters. Notably, it markedly minimized SOC violations and total grid expenses, all while preserving computational agility.

\end{enumerate}

\section*{Acknowledgments}
The authors gratefully acknowledge the Australian Energy Market Operator (AEMO) for providing the essential real-time data utilized in this research.


\section*{Appendix}
\label{sec:appendix}

\begin{algorithm}
\caption{Gradient-based optimization for solving (1)}
\begin{algorithmic}[1]
\Procedure{Optimizer }{$loss_{\text{fn}}$, $var$, $bounds$, $init_{\text{lr}}$, $d_{\text{steps}}$, $d_{\text{rate}}$, $n_{\text{epochs}}$}
\begin{itemize}
\item Initialize \textbf{optimizer} with Adam optimizer, $lr$ = $init_{\text{lr}}$
\item Define \textbf{$lr_{\text{sch}}$} as Exponential Decay with $d_{\text{steps}}$ and $d_{\text{rate}}$
\item For each epoch in $n_{\text{epochs}}$:
    \begin{itemize}
    \item Compute \textbf{loss} using $loss_{\text{fn}}$, and $var$
    \item Compute gradient \textbf{grads} of loss w.r.t. $var$
    \item Update optimizer's $lr$ with \textbf{$lr_{\text{sch}}$} 
    \item Apply gradients to $var$
    \item Project $var$ between lower and upper $bounds$
    \end{itemize}
\item Return $var$
\end{itemize}
\EndProcedure

\Procedure{Main}{}
\begin{itemize}
\item Initialize parameters and variables: $n_{\text{epochs}}$,  $\text{tol}$, $\text{loss}_{\text{prev}}$, $P^{g_{\text{min}}}$,  $P^{g_{\text{max}}}$, $P^{b_{\text{min}}}$, $P^{b_{\text{max}}}$, $\text{batt}_{\text{cap}}$, $P^g_t$, $P^b_t$, $\alpha^{g}_t$, $\alpha^{b}_t$, $\alpha_{\text{opt}}$, $\text{eps}$, $\alpha^{g_{\text{min}}}$, $\alpha^{g_{\text{max}}}$, $\text{SOC}_{\text{init}}$ and
$\text{uncertainty envelope for each state}$. 

\item For each epoch in $n_{\text{epochs}}$:
    \begin{itemize}
    \item Optimize $P_t^g$ and $P_t^b$ using \Call{Optimize}{}
    \item Ensure $P_t^g$ is non-zero during high tariff periods
    \item Compute loss and grads for $\alpha_t^g$ and $\alpha_t^b$
    \item Update $\alpha_t^g$ and $\alpha_t^b$ using grads
    \item Project $\alpha_t^g$ and $\alpha_t^b$ within range [1, 1000]
    \item If convergence criteria are met, break loop
    \end{itemize}
\item Compute \textbf{total cost}, \textbf{SOC violations} and \textbf{SOC target}

\end{itemize}
\EndProcedure
\end{algorithmic}
\end{algorithm}

The loss function is designed according to (1), incorporating constraints as penalties. Despite subsequent adjustments to this function, the core objective of cost minimization remains unchanged as follows:

{\small
\begin{align}
L = &\sum_{t} \alpha_g(t) \cdot \max( P^g_t - P^{g,\text{max}}_t, 0)^2 \nonumber \\
&+ \sum_{t} \alpha_b(t) \cdot \max(P^b_t - P^{b,\text{max}}_t, 0)^2 \nonumber \\
&+ \lambda_{SOC} \cdot \sum_{t} \max(0.2 - SOC_t, 0)^2 \nonumber \\
&+ \lambda_{SOC} \cdot \sum_{t} \max(SOC_t - 0.8, 0)^2 \nonumber \\
&+ \lambda_{balance} \cdot \sum_{t} (P^d_t - P^g_t - P^b_t - P^pv_t)^2 \nonumber \\
&+ \lambda_{\alpha^g_t} \cdot \sum_{t} (\alpha^{g}_t - \alpha^{g'}_t)^2 \nonumber \\
&+ \lambda_{endSOC} \cdot \max(0.8 - SOC_t, 0)^2
\end{align}
}

\noindent where $\alpha^{g'}_t$ is the normalized grid tariff $C^g_t$ mapped to the range of [$\alpha^{g,\text{min}}, \alpha^{g,\text{max}}$]. In our study, normalization is applied to the data presented in Section IV, with power variables divided by 1000 and costs divided by 100. This process ensures an equitable treatment of all variables during algorithm training, facilitates expedited learning and averts potential numerical instability. The primary aim of this analysis is to optimally balance the SOC and grid interactions in the objective function, using weights $\lambda_{\text{SOC}} = 10$, $\lambda_{\text{balance}} = 10000$, $\lambda_{\alpha_g} = 100$, and $\lambda_{\text{endSOC}} = 1000$. These weights ensure high-quality solutions while maintaining feasibility and avoiding constraint violations. The optimization halts when the absolute difference in loss across epochs is below a set tolerance, yielding outputs like $P^g_t$, $P^b_t$, and penalty multipliers $\alpha_g$ and $\alpha_b$. Key metrics such as total electricity cost, SOC violations, final SOC, and optimization duration are computed, with convergence verified every 100 epochs when the loss difference meets the tolerance criteria.

\bibliographystyle{ieeetr}

\bibliography{ref.bib}



\end{document}